\documentclass[12pt]{article}
\usepackage{jheppub}

\usepackage[T1]{fontenc} 
\usepackage[utf8]{inputenc} 
\usepackage{microtype} 
\usepackage{lmodern} 
\usepackage{booktabs} 
\usepackage{mdframed} 
\usepackage{graphicx}

\usepackage[backgroundcolor=white]{todonotes}

\usepackage{hyperref}
 \usepackage{color}
 \definecolor{dark-red}{rgb}{0.4,0.15,0.15}
 \definecolor{dark-blue}{rgb}{0.15,0.15,0.4}
 \definecolor{medium-blue}{rgb}{0,0,0.5}
 \hypersetup{colorlinks, linkcolor={dark-red}, citecolor={dark-blue}, urlcolor={medium-blue}}

\usepackage{amsmath,amsfonts,amssymb}
\usepackage{eucal} 
\usepackage{mleftright} \mleftright
\usepackage{tensor} 
\usepackage{braket} 
\usepackage{bm} 

\usepackage{mathrsfs} 

\providecommand*{\carp}{\phi^{(\Delta,s)}(u,z,\zb)}
\renewcommand*{\carp}{\phi^{(\Delta,s)}(u,z,\zb)}

\newcommand*{\dd}{\mathop{}\!d}

\newcommand*{\scri}{\ensuremath{\mathscr{I}}}

\newcommand*{\zb}{\bar{z}}

\newcommand*{\phib}{\bar{\phi}}
\newcommand*{\sgn}{\textrm{sgn}}
\newcommand*{\veps}{\eta}

\begin{document}
\title{An Embedding Space Approach to Carrollian CFT Correlators for Flat Space Holography}

\author{Jakob Salzer}%
 \emailAdd{jakob.salzer@ulb.be}
 \affiliation{Université Libre de Bruxelles and International Solvay Institutes, ULB-Campus Plaine CP231,
 B-1050 Brussels, Belgium}

\abstract{Carrollian conformal field theories (carrollian CFTs) are natural field theories on null infinity of an asymptotically flat spacetime or, more generally, geometries with conformal carrollian structure. Using a basis transformation, gravitational S-matrix elements can be brought into the form of correlators of a carrollian CFT. Therefore, it has been suggested that carrollian CFTs could provide a co-dimension one dual description to gravity in asymptotically flat spacetimes. In this work, we construct an embedding space formalism for three-dimensional carrollian CFTs and use it to determine two- and three-point correlators. These correlators are fixed by the global subgroup, $\textrm{ISO}(3,1)$, of the carrollian conformal symmetries, i.e., the Bondi--van der Burg--Metzner--Sachs symmetries ($\textrm{BMS}$). The correlators coincide with well-known two- and three-point scattering amplitudes in Minkowski space written with respect to a basis of asymptotic position states.}
\maketitle


\section{Introduction}
\label{sec:introduction}
The duality between quantum gravity on asymptotically Anti-de Sitter (AdS) spacetimes and conformal field theories (CFTs) defined on the boundary of the former \cite{Maldacena:1997re, Klebanov:1999tb, Witten:1998qj} has been the most impressive concrete realization of the holographic principle \cite{Hooft:1993gx,Susskind:1994vu}. Yet this principle is expected to hold regardless of the curvature of the underlying spacetime (cf. \cite{Bousso:2002ju} for a review). Similar to the AdS/CFT correspondence, flat space holography aims to establish a description of quantum gravity in asymptotically flat spacetimes in terms of a lower-dimensional non-gravitational theory such that the holographic principle is made manifest. First attempts in this direction started soon after the discovery of the AdS/CFT correspondence \cite{Polchinski:1999ry,Susskind:1998vk,Giddings:1999jq,deBoer:2003vf}. More recently, the flat space holography program gained additional impetus by emphasizing the fact \cite{Strominger:2013jfa} that asymptotically flat spacetimes exhibit an infinite number of symmetries in contrast to AdS spacetimes (in dimensions greater than three)\footnote{$\textrm{AdS}_3$ is similar to the case of asymptotically flat spacetimes since, in the presence of gravity, the finite-dimensional AdS symmetries are uplifted to two copies of the Virasoro algebra \cite{Brown:1986nw}.}. In the presence of gravity, the Poincaré symmetries of Minkowski space are enhanced to an infinite-dimensional group of asymptotic symmetries called Bondi--van der Burg--Metzner--Sachs (BMS) group \cite{Bondi:1962px,Sachs:1962zza,Barnich:2010eb,Barnich:2011mi,Troessaert:2017jcm,Henneaux:2018cst,Henneaux:2018hdj}. These symmetries are comprised of an infinite number of abelian supertranslations together with an infinite number of superrotations that form two copies of the Witt algebra.\footnote{This notion of BMS symmetries is usually known as ``extended BMS'' symmetry going back to \cite{Barnich:2010eb,Barnich:2011mi}. In another proposed extension \cite{Campiglia:2014yka}, known as ``generalized BMS'', superrotations are taken to be elements of $\textrm{Diff}(S^2)$. In the following, we will always refer to the notion of ``extended BMS''.} A dual theory to gravity in asymptotically flat spacetimes must realize these symmetries.

One very basic feature of the $\textrm{AdS}_{d+1}/\textrm{CFT}_d$ correspondence is the fact that the symmetry group $\textrm{SO}(d,2)$ acts geometrically on both sides of the correspondence: on the one hand as the Killing symmetries of the vacuum on the gravity side and on the other hand as the conformal symmetries of $d$-dimensional Minkowski space \cite{Witten:1998qj}. This can be rephrased as the existence of two homogeneous spaces of different dimension for the group $\textrm{SO}(d,2)$, one of which allows for the definition of a Lorentzian metric of constant negative curvature, the other one only for a conformal class of metrics. Mirroring this argument in the case of four-dimensional asymptotically flat gravity, one should therefore look for another, lower-dimensional homogeneous space of the Poincaré group $\textrm{ISO}(3,1)$, i.e., the symmetries of four-dimensional Minkowski space. Interestingly, there exists only one such space of dimension three \cite{Figueroa-OFarrill:2021sxz}. It is not possible to define a metric on this space, rather one finds a \emph{conformal carrollian structure}: an equivalence class of \emph{carrollian structures}, given by a pair $(q_{ab},n^a)$ of vector $n^a$ and degenerate metric $q_{ab}$ with $q_{ab}n^a=0$, that are identified if differing only by scale, $(q_{ab},n^a) \sim (\lambda^2 q_{ab},\lambda^{-1}n^a)$. The symmetries preserving this structure
\begin{equation}
  \label{eq:57}
  \mathcal{L}_\xi n^{a}=-\lambda n^{a},\qquad \mathcal{L}_\xi q_{ab}=2\lambda q_{ab}\,,
\end{equation}
are nothing but the above-mentioned infinite-dimensional BMS symmetries \cite{Geroch1977,Ashtekar:1981hw,Duval:2014uva}. In analogy to the case of AdS/CFT, one might then expect that a three-dimensional \emph{carrollian CFT}, i.e., a three-dimensional BMS-invariant field theory, as a dual of gravity in asymptotically flat spacetimes. This has been dubbed the carrollian approach to flat space holography \cite{Banerjee:2018gce,Donnay:2022aba,Bagchi:2022emh,Donnay:2022wvx,Saha:2023hsl}, which has been previously explored in the context of three-dimensional gravity; see, e.g., \cite{Bagchi:2010zz,Bagchi:2012xr, Bagchi:2012cy,Barnich:2013yka,Bagchi:2014iea,Bagchi:2015wna,Hartong:2015usd,Ciambelli:2018wre,Merbis:2019wgk}.

A complementary approach to flat space holography, known as the celestial CFT program \cite{Pasterski:2016qvg,Pasterski:2017ylz} aims to describe the holographic dual to gravity in asymptotically flat spacetimes as a two-dimensional CFT. This approach has led to a number of fascinating results: among them the derivation of conformal soft theorems from BMS invariance \cite{Donnay:2018neh,Fan:2019emx,Puhm:2019zbl,Donnay:2020guq}, a celestial OPE from collinear limits \cite{Fotopoulos:2019tpe,Pate:2019lpp,Banerjee:2020kaa,Adamo:2021zpw} and evidence for a $w_{1+\infty}$ symmetry of (tree-level) scattering in gravity \cite{Guevara:2021abz,Strominger:2021mtt,Ball:2021tmb,Freidel:2021ytz}.

The most natural observable for a theory in asymptotically flat spacetimes is the S-matrix which can be regarded as holographic by its very definition as a function of boundary data. Consequently, both the celestial and the carrollian approach to flat space holography aim to compute the S-matrix in terms of correlators of the respective theories. In the celestial CFT approach, a correlator of primary states of the two-dimensional CFT is claimed to compute S-matrix elements in a basis of asymptotic boost eigenstates \cite{Pasterski:2016qvg,Pasterski:2017ylz,Pasterski:2017kqt}. In the carrollian CFT approach, a correlator of primary fields in the three-dimensional carrollian CFT computes S-matrix elements in a basis of asymptotic position eigenstates.

Both approaches are taylored to different aspects of gravity in asymptotically flat spacetimes and both face some difficulties. In carrollian CFTs, all symmetries are realized as vector fields on the underlying three-dimensional conformal carrollian manifold, as was discussed above.\footnote{In contrast, in celestial CFTs translations act by shifting the conformal weights. We can understand this along the lines of the argument given above. The two-sphere endowed with a conformal metric is no homogeneous space for the Poincaré group but only for the Lorentz group. Consequently, translations cannot be realized geometrically.} The obvious drawback in the conformal carrollian approach is the lack of concrete examples of interacting carrollian CFTs, cf. \cite{Bagchi:2019xfx,Barnich:2013yka,Barnich:2022bni,Banerjee:2020qjj,Chen:2023pqf,Mehra:2023rmm} for some examples, and, relatedly, our very incomplete understanding of these theories.

In this work we want to take some steps towards a better understanding of carrollian CFTs by determining the form of two- and three-point functions of a generic carrollian CFT. In a conventional CFT, two- and three-point functions of quasi-primary fields are fixed by global conformal invariance. It is therefore to be expected that the form of two- and three-point functions in carrollian CFTs is also determined by the global subgroup of $\textrm{BMS}$, i.e., the Poincaré group. This has been demonstrated in \cite{Bagchi:2022emh,Donnay:2022wvx} for the case of the two-point function by explicitly solving the corresponding Ward identities. The authors found two branches: the first one takes the form of the two-point correlator of a two-dimensional CFT, while the second branch is distributionally valued. It is this second branch that computes the two-point scattering amplitude in an asymptotic position basis. We will find similarly that the three-point function also has two different branches: it can either take the form of a three-point correlator in a two-dimensional CFT or it can be distributionally valued. This latter branch coincides with the corresponding three-particle scattering amplitude around Minkowski space written in position space.

In order to compute these correlators, we will develop a carrollian version of the embedding space formalism for conventional CFTs. This formalism uses the fact that the conformal group in $d$-dimensions, $\textrm{SO}(d,2)$, can be seen as the analogue of the Lorentz group in a $(d+2)$-dimensional space of signature $(d,2)$. Objects constructed using the metric in $(d+2)$ dimensions and suitably projected down to the $d$-dimensional space are then guaranteed to be conformally invariant in $d$ dimensions. This relation between conformal invariance in $d$ dimensions and Lorentz invariance in $d+2$ dimensions was first realized by Dirac \cite{Dirac:1936fq} and used in the early CFT literature \cite{Mack:1969rr,Boulware:1970ty,Ferrara:1973yt}; see also \cite{Weinberg:2010fx} for more references to earlier literature. Its modern form, that is essential to many applications in AdS/CFT, was developed in \cite{Cornalba:2009ax,Costa:2011mg,Simmons-Duffin:2012juh,Costa:2014rya,Elkhidir:2014woa}; in the following we will rely on the presentation given in \cite{Costa:2011mg}. As we will see in detail below, the main ingredient to produce conformal carrollian correlators is the introduction of a preferred null vector in the embedding space that explicitly breaks $\textrm{SO}(4,2)$ down to $\textrm{ISO}(3,1)$, and a subsequent limiting procedure that pushes the correlators to a preferred null surface.

The construction of this carrollian embedding space formalism will be the main result of the present work, with the construction of low-point correlators only a first application. We expect that this formalism will be useful for the study of carrollian CFT; a selection of immediate applications is presented in the discussion in Section \ref{sec:discussion}.

This article is organized as follows. In Section \ref{sec:car2pt}, we will introduce the six-dimensional embedding space and show how to define conformal carrollian quasi-primaries on this embedding space. In Section \ref{eq:carcor2pt}, we will use the embedding space to reproduce the two-point conformal carrollian correlator of a carrollian scalar field, previously found by an explicit solution of the Ward identities. In Section \ref{sec:spin2ptfun}, we generalize the discussion to discuss the two-point correlator of spinning fields. In Section \ref{sec:scalar-fields} and Section \ref{sec:3ptfun}, we use the embedding space approach to compute the three-point correlator of carrollian scalar and spin one fields and indeed reproduce the three-point amplitudes of these fields around Minkowski space. We conclude with a discussion in Section \ref{sec:discussion}. The appendices contain a summary of our conventions in Appendix \ref{sec:embcoord}, a discussion of the relation between the symmetries of the embedding space and the global conformal carrollian symmetries in Appendix \ref{sec:syminembed}, and two- and three-point scattering amplitudes in Minkowski space written in a basis of asymptotic position eigenstates in Appendix \ref{sec:scatt-ampl-moment}.

\section{Conformal Carrollian fields from the embedding space}
\label{sec:car2pt}

In this section, we want to develop an embedding space approach for conformal carrollian fields. By this we mean a formalism that allows us to write down explicitly $\textrm{ISO}(3,1)$-invariant quantities in a six-dimensional space. Subsequently, we will project this quantities to a three-dimensional hypersurface with a conformal carrollian structure. In this way, we obtain explicitly Poincaré-invariant quantities on the conformal carrollian manifold. To set the stage, we will first briefly review the embedding space formalism for four-dimensional CFTs and then discuss the necessary modifications to account for conformal carrollian theories.

\subsection{The embedding space formalism for CFTs}
\label{sec:embedforCFTs}
The conformal group in four dimensions is $\textrm{SO}(4,2)$. To discuss
the constraints imposed by conformal invariance on CFT observables, in particular on correlators, it is very convenient to formulate the problem in a space where the conformal group acts linearly. The most natural choice for such a space is $\mathbb{R}^{4,2}$, the so-called embedding space. Treatments can be found
in \cite{Cornalba:2009ax,Weinberg:2010fx, Costa:2011mg,Simmons-Duffin:2012juh,Costa:2014rya,Elkhidir:2014woa} but the methods go back to
Dirac \cite{Dirac:1936fq}.

We will choose coordinates
\begin{equation}
  \label{eq:16}
  X^A=(X^\mu,X^+,X^-)^A\,, \qquad \mu=0,\dots, 3
\end{equation}
 on $\mathbb{R}^{4,2}$
with metric
\begin{equation}
  \label{eq:31}
  \dd s^2=g_{AB}\dd X^A \dd X^B=\eta_{\mu\nu}\dd X^\mu\dd X^\nu+2 \dd X^+ \dd X^-\,,
\end{equation}
where $\eta_{\mu\nu}$ is the four-dimensional Minkowski metric of signature $(-+++)$, so that the metric $g_{AB}$ is of signature $(-++++-)$.
The usual embedding space for four-dimensional Lorentzian CFTs is
obtained by considering the hypercone $X^Ag_{AB}X^B\equiv X\cdot X=0$
and going over to the projective space by identifying
$X^A \sim \lambda X^A, \lambda>0$. We will parametrize points on the surface
$X\cdot X=0$ as
\begin{equation}
  \label{eq:82a}
 X^{A}=\sigma(x^\mu,-x^2/2,1)^A\equiv \sigma X^{A}_x\,,\qquad \sigma>0.
\end{equation}
It is conventional to use the scaling freedom to set $\sigma=1$. However, we will refrain from doing so since we will be interested precisely in the regions where $\sigma \to 0$.

The Killing vectors that preserve the metric \eqref{eq:31} are given by
\begin{equation}
  \label{eq:9}
  M_{AB}=X_{A}\partial_B-X_B\partial_A\,,
\end{equation}
and they generate the algebra $\mathfrak{so}(4,2)$. Note that these Killing vectors keep invariant the surface $X^2=0$.

Let $\Phi^{\Delta}_{A_1...A_s}(X)$ be a rank $s$ tensor field
defined on the cone $X^2=0$,
with the requirements
\begin{subequations}
    \label{eq:homo}
\begin{align}
  \Phi^{\Delta}_{A_1...A_s}(\lambda X)&=\lambda^{-\Delta}\Phi_{A_1...A_s}(X)\Leftrightarrow X^A\partial_A\Phi^{\Delta}_{A_1...A_s}(X)=-\Delta \Phi^{\Delta}_{A_1...A_s}(X) \\
  \label{eq:trans}
  X^{A_i}\Phi^{\Delta}_{A_1..A_i..A_s}&=0 \qquad \forall i
\end{align}
\end{subequations}
 The projection
\begin{equation}
  \label{eq:projectrel}
\phi_{\mu_1...\mu_s}(x)=\sigma^{\Delta-s}\frac{\partial X_x^{A_1}}{\partial x^{\mu_1}}...\frac{\partial X_x^{A_s}}{\partial x^{\mu_s}}\Phi^{\Delta}_{A_1...A_s}(X_x)
\end{equation}
defines a four-dimensional tensor field. Note that we can add terms proportional to $X_A$ to $\Phi^{\Delta}_{A_1...A_s}$
without changing $\phi_{\mu_1\ldots \mu_s}$ since
\begin{equation}
  \label{eq:10}
  \partial_\mu X^A X_A=0\,
\end{equation}
on the cone. Fields $\Phi^{\Delta}_{A_1\ldots A_s}$ differing by such terms can therefore be regarded as gauge-equivalent. This reduces the number of independent components of the field $\Phi^{\Delta}_{A_1\ldots A_s}$ from $6^s$ to $4^s$, the correct number for a four-dimensional field of rank $s$.

It is now a straightforward computation (cf., e.g., \cite{Weinberg:2010fx}) to show that the four-dimensional field defined in \eqref{eq:projectrel} transforms like a primary field under the action of the Killing symmetries on $\Phi^{\Delta}_{A_1...A_s}(X_x)$. The a priori arbitrary parameter $\Delta$ becomes the conformal dimension of the CFT primary.

This formalism allows a straightforward determination of the constraints on correlation functions imposed by conformal invariance. The two-point correlator of two scalar fields in embedding space must be of the form
\begin{equation}
  \label{eq:12a}
  \langle \Phi^{\Delta_1}(X_1)\Phi^{\Delta_2}(X_2)\rangle=f(X_{12})\,,
\end{equation}
where $X_{12}=X^A_1g_{AB}X^B_2$. This is the only scalar we can build from the metric and the two points $X^A_1,X^A_2$. The scaling condition \eqref{eq:homo} now restricts this function further to be of the form
\begin{equation}
  \label{eq:12}
  \langle \Phi^{\Delta_1}(X_1)\Phi^{\Delta_2}(X_2)\rangle=\frac{C_2}{X^\Delta_{12}}\qquad \Delta=\Delta_1=\Delta_2\,.
\end{equation}
The correlator in the physical spacetimes is now obtained by using the parametrization \eqref{eq:82} and the relation \eqref{eq:projectrel} so that one finds
\begin{equation}
  \label{eq:58}
  \langle \phi^{\Delta_1}(x_1)\phi^{\Delta_2}(x_2)\rangle=\frac{C'_2}{|x_1-x_2|^{2\Delta}}\qquad\Delta=\Delta_1=\Delta_2\,.
\end{equation}
Note that we need to include an $i\epsilon$ prescription in \eqref{eq:12} and \eqref{eq:58} in order to circumvent the respective poles.
For the case of a three-point function of scalar fields, one finds using the same arguments in embedding space
\begin{equation}
  \label{eq:87}
  \langle \Phi^{\Delta_1}(X_1)\Phi^{\Delta_2}(X_2)\Phi^{\Delta_3}(X_3)\rangle=C_3\prod_{\circlearrowleft ijk}X^{\frac{\Delta_i+\Delta_j-\Delta_k}{2}}_{ij}\,,
\end{equation}
which directly leads to the correct physical three-point function
\begin{equation}
  \label{eq:58a}
  \langle \phi^{\Delta_1}(x_1)\phi^{\Delta_2}(x_2)\phi^{\Delta_3}(x_3)\rangle=\frac{C'_3}{|x_{12}|^{\Delta_{12}-\Delta_3}|x_{23}|^{\Delta_{23}-\Delta_1}|x_{31}|^{\Delta_{31}-\Delta_2}}\,,
\end{equation}
where $\Delta_{ij}=\Delta_i+\Delta_j$. Two and three-point functions are thus completely determined by conformal symmetry up to normalization.

The case of spinning fields proceeds analogously. The difficulty lies in determining the correct tensor structures that obey the transversality condition in \eqref{eq:homo}. We will discuss this in some detail directly in the carrollian case; for a thorough treatment in the CFT case see \cite{Costa:2011mg,Elkhidir:2014woa}.

Although we discussed here the case of four-dimensional CFTs, a similar strategy works for CFTs in any dimension, in particular also for two-dimensional CFTs. In this case, conformal symmetries are infinite-dimensional and the vector fields \eqref{eq:9} generate only the global subalgebra $\mathfrak{so}(2,2)$. Consequently, fields defined like \eqref{eq:projectrel} from the embedding space field will transform like quasi-primary fields. The form of two- and three-point functions for quasi-primary fields is again completely specified by the global subalgebra $\mathfrak{so}(2,2)$. We mention this explicitly, since the situation for carrollian CFTs is somewhat comparable, with the BMS algebra playing the role of the local conformal symmetries and the Poincaré algebra playing the role of the global subalgebra that fixes two- and three-point correlators.

Having briefly reviewed the embedding space for CFTs we will now turn to describe the necessary ingredients to adapt this set-up for
carrollian CFTs.
\subsection{Null infinity in the embedding space}
\label{sec:null-infin-embedd}
In the following, we will find it convenient to introduce a particular choice of the coordinates $x^\mu$ in \eqref{eq:82a}. We will from now on parametrize the points $X^A$ on the surface $X^2=0$ in the embedding space $\mathbb{R}^{4,2}$ as
\begin{equation}
  \label{eq:82}
X^{A}=\sigma(x^\mu,-x^2/2,1)^A\equiv \sigma X^{A}_x\,,\qquad x^\mu= u n^\mu+r q^\mu(z,\zb)\,, \qquad \sigma>0,
\end{equation}
where $q^\mu(z,\zb)$ is a null vector pointing to a point $(z,\zb)$ on the stereographically projected two-sphere and $n^\mu$ is another fixed null-vector; see Appendix \ref{sec:embcoord} for our coordinate conventions.

In the previous section, we showed how the embedding space formalism can be used to write manifestly $\textrm{SO}(4,2)$-invariant correlators. In the following, however, we
will be interested in Poincaré-invariant quantities, i.e., invariant under the subgroup $\textrm{ISO}(3,1)$ of $\textrm{SO}(4,2)$. Let us therefore introduce a distinguished null vector
$I^A$, $I^Ag_{AB}I^B=0$, with coordinates
\begin{equation}
  \label{eq:50}
  I^{A}=(0^\mu,1,0)\,.
\end{equation}
The subgroup of $\textrm{SO}(4,2)$ that keeps this null vector
invariant in the sense $M\indices{^A_B} I^B=I^A$ is $ISO(3,1)$.\footnote{This is analogue to the subgroup $\textrm{ISO}(2)$ of the Lorentz group $\textrm{SO}(3,1)$ that keeps null vectors in Minkowski space invariant.} $I^A$ thus breaks the conformal symmetry in the
embedding space down to Poincaré symmetries. In related, more
general context, this object is known as infinity tractor or infinity
twistor cf., e.g., \cite{Penrose:1972ia,Penrose:1985bww,Hughston:1983jep,Curry:2014yoa}, and \cite{Herfray:2021xyp,Herfray:2021qmp,Figueroa-OFarrill:2021sxz} for more recent works with explicit applications to the geometry of null infinity. We will return briefly to this in the discussion.

Holding all other coordinates fixed in the parametrization~\eqref{eq:82}, setting $\sigma=0$ clearly leads
to $X=0$. However, if we let both $\sigma\rightarrow 0$ and
$r \rightarrow \infty$ keeping $r\sigma$ finite or, equivalently, use part of the scaling freedom to fix $\sigma$ to $|r|$, we obtain
\begin{equation}
  \label{eq:47}
  X^{A}_{\scri}=(q^\mu,u,0) \sim \lambda X^{A}_{\scri} \, .
\end{equation}
The last identification follows since we still have the freedom of
rescaling $\sigma \rightarrow \lambda \sigma$ by a positive function $\lambda$.
This space is a homogeneous
space of the Poincaré group with conformal carrollian structure given by
\begin{equation}
  \label{eq:confcarstructure}
  (q=2 \dd z \dd \zb,n=\partial_u)\sim (\lambda^2q,\lambda^{-1}n)\,.
\end{equation}
We can identify this space with null infinity.
In the following, we will
fix the freedom of rescaling by setting $r\sigma=1$. The induced
degenerate metric on $\scri$ is then the flat metric
$\dd s^2|_{\scri}=2\dd z \dd \zb$. If one chose to fix
$r\sigma=(1+z\zb)^{-1}$ on the other hand, the induced metric on the boundary would
correspond to the metric on the two-sphere. While our results are also
applicable in the latter case, we will stick, for brevity and easier comparison to other approaches in the literature, to the
former representative.

In Minkowski space, the flat Bondi coordinates \eqref{eq:flatbondi}
cover both future $(r\rightarrow +\infty)$ and past
$(r\rightarrow -\infty)$ null infinity. We can therefore set $r\sigma=-1$ and mirror the
limit leading to \eqref{eq:47} in order to obtain past null infinity.

\subsection{Defining conformal carrollian primaries}
\label{eq:correlators}

A carrollian CFT is defined as a theory that naturally couples to a
background with conformal carrollian structure \eqref{eq:confcarstructure} \cite{Bagchi:2019xfx,Donnay:2022aba,Bagchi:2022emh,Baiguera:2022lsw,Chen:2021xkw}. This background
structure is left-invariant by conformal carrollian symmetries, i.e.,
the extended four-dimensional BMS transformations
\cite{Barnich:2010eb,Barnich:2011mi,Duval:2014lpa}. In the coordinates chosen in \eqref{eq:confcarstructure}, these vector fields are given by
\begin{equation}
  \label{eq:BMSvectors}
\xi=\left(\mathcal{T}(z,\zb)+\frac{u}{2}(\partial Y+\bar{\partial}\bar{Y})\right)\partial_u+Y\partial+\bar{Y}\bar{\partial}\,.
\end{equation}
Here, $Y(z)$ $(\bar{Y}(\zb))$ is a (anti-)holomorphic local conformal
transformation on the (punctured) plane parametrizing superrotations,
while $\mathcal{T}(z,\zb)$ parametrizes supertranslations. The global
subalgebra, i.e., the Poincaré algebra, is generated by
$\mathcal{T}=1,z,\zb,z \zb$, $Y(z)=1,z,z^2$ and its conjugate.
A conformal carrollian (quasi-)primary
field of weight $\Delta$ and helicity $+s$$[-s]$ denoted by $\phi^{(\Delta, s)}_{\underbrace{\scriptstyle z\ldots z}_{\scriptstyle s}}(u,z,\zb)$ $[\phi^{(\Delta, s)}_{\underbrace{\scriptstyle \zb\ldots \zb}_{\scriptstyle s}}(u,z,\zb)]$ is then defined as a field on a conformal
carrollian manifold transforming as
\begin{subequations}
    \label{eq:ccprimary}
\begin{align}
  \delta_\xi\phi^{(\Delta,s)}_{z\ldots z}(u,z,\zb)&=\bigg[\left(\mathcal{T}(z,\zb)+\frac{u}{2}(\partial Y+\bar{\partial}\bar{Y})\right)\partial_u +Y\partial+\frac{\Delta+ s}{2}\partial Y\\
                                                    &\qquad \qquad+\bar{Y}\bar{\partial}+\frac{\Delta- s}{2}\bar{\partial}\bar{Y} \bigg] \phi^{(\Delta, s)}_{z \ldots z}(u,z,\zb),\nonumber\\
 \delta_\xi\phi^{(\Delta,s)}_{\zb\ldots \zb}(u,z,\zb)&=\bigg[\left(\mathcal{T}(z,\zb)+\frac{u}{2}(\partial Y+\bar{\partial}\bar{Y})\right)\partial_u +Y\partial+\frac{\Delta- s}{2}\partial Y\\
                                                    &\qquad \qquad+\bar{Y}\bar{\partial}+\frac{\Delta+s}{2}\bar{\partial}\bar{Y} \bigg] \phi^{(\Delta, s)}_{\zb \ldots \zb}(u,z,\zb).\nonumber
\end{align}
\end{subequations}
 If the transformation \eqref{eq:ccprimary} is valid only for $\mathcal{T},Y,\bar{Y}$ being part of the Poincaré subalgebra, the field is called quasi-primary. In what follows, we will be interested almost exclusively in quasi-primary fields.

We are now in the position to relate conformal carrollian quasi-primary fields
$\carp$ to fields on the embedding
space. For this purpose, we define fields $\Phi^{\Delta}_{A_1\dots A_s}(X)$ on
the hypercone $X^{2} = 0$ in the $\mathbb{R}^{4,2}$ embedding space
transforming as tensors under $\textrm{SO}(4,2)$. In particular, these
fields transform as tensors under the Poincaré group, i.e., the
subgroup of $\textrm{SO}(4,2)$ keeping $I^A$ fixed. We furthermore
require these fields to behave homogeneously under a rescaling of the coordinates
\begin{equation}
  \label{eq:homogscale}
\Phi^{\Delta}_{A_1\dots A_s}(\lambda X)=\lambda^{-\Delta}\Phi^{\Delta}_{A_1\dots A_s}(X)\,,
\end{equation}
where $\Delta$ is a priori undetermined.
The field $\Phi^{\Delta}_{A_1\ldots A_s}$ must obey the conditions
\begin{align}
  \label{eq:transverscon}
  X^A\Phi^{\Delta}_{A_1\dots A_s}&=0 &  I^A\Phi^{\Delta}_{A_1\dots A_s}&=0\,.
\end{align}
We will further require $\Phi^{\Delta}_{A_1\ldots A_s}$ to be symmetric and traceless with respect to $g^{AB}$. We saw above that the limit $\sigma\to 0$ with $r\sigma=\pm 1$ in the coordinates \eqref{eq:82} takes one $\scri^\pm$. Thus, we find that the conformal carrollian field $\carp$ defined on $\scri^\pm$ can be related to an embedding space field $\Phi^{\Delta}_{A_1\ldots A_s}$ of weight $\Delta$ as
\begin{subequations}
    \label{eq:centralchain}
\begin{align}
  \phi^{(\Delta,s)}_{z\ldots z}&=\lim_{\sigma \to 0}(\partial_z X^{A_1}\ldots \partial_z X^{A_s}\Phi^{\Delta}_{A_1\ldots A_s}(X))_{r\sigma=\veps}\\
  \phi^{(\Delta,s)}_{\zb\ldots \zb}&=\lim_{\sigma \to 0}(\partial_{\zb} X^{A_1}\ldots \partial_{\zb} X^{A_s}\Phi^{\Delta}_{A_1\ldots A_s}(X))_{r\sigma=\veps}\,.
\end{align}
\end{subequations}
with $\veps=\pm 1$, depending on whether the field is defined on future null infinity $(+1)$ or past null infinity $(-1)$. Indeed, it can be shown that the transformations \eqref{eq:9} of $\Phi^{\Delta}_{A_1\ldots A_s}(X)$ preserving $I^{A}$ lead to the appropriate transformations for a conformal carrollian quasi-primary field (cf. Appendix \ref{sec:syminembed}). To summarize: \emph{A
  field $\Phi^{\Delta}_{A_1\dots A_s}(X)$ defined on the hypercone $X^{2}=0$ in
  embedding space and of scaling dimension $\Delta$ that obeys conditions
  \eqref{eq:homogscale} and \eqref{eq:transverscon} yields, contracted with polarization vectors
  $\partial_zX^A$, a conformal carrollian field
  $\phi^{(\Delta,s)}_{z \ldots z}(u,z,\zb)$ when evaluated on the surface
  $r=\veps \sigma^{-1}$ in the limit $\sigma\to 0$. The quantity
  $\veps=\pm 1$ determines whether the field is defined on future
  $(+1)$ or past $(-1)$ null infinity.}

The above definition allows the description of conformal carrollian field of any weight $\Delta$ from the embedding space. In the approach advocated in \cite{Donnay:2022aba, Donnay:2022wvx}, only conformal carrollian fields with weight $\Delta=1$ are argued to be relevant in the dual description of (gravitational) scattering in asymptotically flat spacetimes. The reason is that these fields can be explicitly related to the free radiative data in the asymptotic expansion of massless fields in asymptotically flat spacetimes. We can also argue for the relation \eqref{eq:centralchain} from this perspective.

Starting from a
four-dimensional massless bulk field $\phi^{(s)}_{\mu_1\dots \mu_s}$,
assumed to be symmetric and to satisfy the De Donder traceless gauge
condition
\begin{equation}
  \label{eq:Dedonder}
  \eta^{\mu\nu}\phi^{(s)}_{\mu\nu\mu_3\dots \mu_s}=0\qquad\partial^{\mu}\phi^{(s)}_{\mu\mu_2\dots \mu_s}=0 
\end{equation}
one defines carrollian boundary fields
$\bar{\phi}^{(s)}_{z\dots z}(u,z,\zb)$
$[\bar{\phi}^{(s)}_{\zb\dots \zb}(u,z,\zb)]$ of spin $s$ and positive
[negative] helicity as the following limit
\begin{align}
    \label{eq:Donnaydef}
  \bar{\phi}^{(s)}_{z\dots z}(u,z,\zb) \otimes^s(\dd z)+  \bar{\phi}^{(s)}_{\zb\dots \zb}(u,z,\zb) \otimes^s(\dd \zb)\nonumber\\
  \equiv \lim_{\Delta \to 1}\lim_{r\to +\infty}\iota(r^{\Delta-s}\phi^{(s)}_{\mu_1\dots \mu_s}(x^{\mu}) \dd x^{\mu_1}\otimes\dots \otimes \dd x^{\mu_s})\,,
\end{align}
where $\iota$ denotes the pull-back to the boundary. Defining the
polarization vectors
\begin{equation}
  \label{eq:defpol}
  \veps^\mu(z,\zb)=\partial_zq^\mu,\qquad \bar{\veps}^\mu(z,\zb)=\partial_{\zb}q^\mu\,,
\end{equation}
we can write \eqref{eq:Donnaydef} as
\begin{equation}
  \label{eq:Donnaydefmod}
   \bar{\phi}^{(s)}_{z\dots z}(u,z,\zb)=\lim_{\Delta \to 1}\lim_{r\to \infty}(r^\Delta \phi^{(s)}_{\mu_1\dots \mu_s}\veps^{\mu_1}\dots \veps^{\mu_s})\,
 \end{equation}
and similar for the opposite helicity.
It can be shown that this prescription fixes $\Delta$ in \eqref{eq:ccprimary} to the value $\Delta=1$ \cite{Donnay:2022aba,Donnay:2022wvx}. Starting from a symmetric, traceless embedding space field $\Phi^{\Delta}_{A_1\ldots A_s}$ of weight $\Delta=1$ we can define a four-dimensional bulk field $\phi^{(s)}_{\mu_1\dots \mu_s}$ using \eqref{eq:projectrel}. Tracelessness of $\Phi^{\Delta}_{A_1\ldots A_s}$ with respect to $g_{AB}$ implies tracelessness of $\phi^{(s)}_{\mu_1\dots \mu_s}$ with respect to $\eta_{\mu\nu}$ so that this field is indeed traceless and symmetric. It remains to implement the De Donder gauge condition \eqref{eq:Dedonder} on the embedding space field. This translates to
\begin{equation}
  \label{eq:DeDondersix}
\partial^\mu\phi^{(s)}_{\mu\mu_2\dots \mu_s}(x^\mu)=\sigma^{-s}\frac{\partial X^{A_2}}{\partial x^{\mu_2}}\dots \frac{\partial X^{A_s}}{\partial x^{\mu_s}}R_{A_2\dots A_s}(X),
\end{equation}
where
\begin{equation}
  \label{eq:defdedon}
  R_{A_2\dots A_s}(X)
  =(I\cdot X)^{-1}
  \left(
    \partial^{A_1}-\frac{(I\cdot \partial)}{(X\cdot I)}X^{A_1}-(l+2-\Delta)I^{A_1}
  \right)
  \Phi^{\Delta}_{A_1\dots A_s}(X)\,,\qquad \Delta=1
\end{equation}
For the gauge condition to hold, we have to require $R_{A_2\dots A_s}(X)=0$, up to terms vanishing when contracted with $\partial_\mu X^A$. The proof of both of the latter two statements
is given in  \cite{Costa:2011mg} with slight adaptions to the present
set-up. In a usual CFT, the operator \eqref{eq:defdedon} is not conformally invariant due to the last term that depends explicitly on $I^A$. The condition $\partial^\mu\phi^{(s)}_{\mu\mu_2\dots \mu_s}(x^\mu)=0$ can only be imposed consistently if the field has conformal dimension such that the last term vanishes. In the present set-up, however, there is no such restriction since $I^A$ is part of the additional structure that is available.

We are now able to directly relate fields in embedding space
$\Phi^{\Delta}_{A_1\dots A_s}(X)$ of scaling dimension $\Delta=1$ to conformal
carrollian fields $\bar{\phi}^{(s)}_{z\dots z}(u,z,\zb)$. In the
coordinates designated in \eqref{eq:82}, set $r^{-1}=\sigma$. Note, that Poincaré transformations preserve
surfaces of constant $\sigma$. We have now
\begin{subequations}
    \label{eq:centralchain2b}
\begin{align}
  &\bar{\phi}^{(s)}_{z\dots z}(u,z,\zb)=\lim_{\Delta \to 1}\lim_{r\to \infty}(r^\Delta \phi_{\mu_1\dots \mu_s}\veps^{\mu_1}\dots \veps^{\mu_s})=\lim_{\Delta \to 1}\lim_{\sigma\to 0}(\sigma^{\Delta} \phi^{(s)}_{\mu_1\dots \mu_s}\veps^{\mu_1}\dots \veps^{\mu_s})|_{r\sigma=1}\nonumber\\
  &=\lim_{\Delta\to 1}\lim_{\sigma \to 0}(\sigma^{\Delta-1-s}\veps^{\mu_1}\frac{\partial X^{A_1}}{\partial x^{\mu_1}}\dots \veps^{\mu_s}\frac{\partial X^{A_s}}{\partial x^{\mu_s}}\Phi^{\Delta}_{A_1\dots A_s}(X))|_{r\sigma=1}\nonumber\\
  &=\lim_{\Delta\to 1}\lim_{\sigma \to 0}\left(\partial_z X^{A_1}\dots \partial_z X^{A_s}\Phi^{\Delta}_{A_1\dots A_s}(X)|_{r\sigma=1}\right)\,
\end{align}
\end{subequations}
and similarly for the other helicity. Here, we subsequently used \eqref{eq:Donnaydefmod}, \eqref{eq:projectrel} and explicitly
evaluated the derivative using \eqref{eq:82}. We see that this coincides with our definition \eqref{eq:centralchain}. In the following, we will find it convenient to work with the more general definition \eqref{eq:centralchain} and take the limit $\Delta\to 1$ on our expressions in the very end. We will see that in this limit many of the correlators diverge so that this provides a natural regularization of the expressions.

\subsection{The S-matrix as a conformal carrollian correlator}
\label{sec:Smatrix}
It has been suggested in \cite{Donnay:2022wvx,Bagchi:2022emh,Banerjee:2018gce} that the S-matrix for massless fields in Minkowski space can be brought into the form of a conformal carrollian correlator after a suitable basis transform. We will refer to these works for further details. Here we just state the results for completeness.

Parametrizing the momentum of a massless field as $p^\mu=\omega q^\mu$, \cite{Donnay:2022wvx} define the position space S-matrix of $n$ outgoing and $N-n$ ingoing fields as
\begin{align}
  \label{eq:65}
  \mathcal{C}(u_1,z_1,\zb_1;\ldots, u_n,z_n,\zb_n)=
  \frac{1}{(2\pi)^N}\prod^n_{k=1}\int^\infty_0\dd \omega_k e^{-\veps_k i\omega_k u_k}\mathcal{A}_N(p_1;\ldots;p_N)\,,
\end{align}
where $\veps_k=\pm 1$ denotes whether the particle is ingoing $(-1)$ or outgoing $(+1)$. The carrollian fields $\bar{\phi}^{(+s)}_{z\dots z}(u,z,\zb),\bar{\phi}^{(s)}_{\zb\dots \zb}(u,z,\zb)$ are related by a Fourier transform to the usual creation/annihilation operators of asymptotic plane wave states \cite{He:2014laa,Gonzo:2020xza,Donnay:2022sdg}. The left hand side of \eqref{eq:65} can therefore be regarded as a correlator of the conformal carrollian fields $\bar{\phi}^{(s)}_{z\dots z}(u,z,\zb),\bar{\phi}^{(s)}_{\zb\dots \zb}(u,z,\zb)$.

In contrast, in \cite{Bagchi:2022emh,Banerjee:2018gce} it was suggested that the relation between correlators of a conformal carrollian theory and the S-matrix in momentum space is given by the modified Mellin transform
\begin{align}
  \label{eq:88}
 \tilde{\mathcal{C}}(u_1,z_1,\zb_1,\Delta_1;\ldots, u_n,z_n,\zb_n,\Delta_n)=
  \prod^n_{k=1}\int^\infty_0\dd \omega_k \omega^{\Delta_i-1}e^{-\veps_k i\omega_k u_k}\mathcal{A}_N(p_1;\ldots;p_N)\,.
\end{align}
It can be shown that the left hand side transforms indeed as a conformal carrollian correlator. In this case, however, the relation between the carrollian fields and the usual  creation/annihilation operators of asymptotic plane wave states is not so obvious.

Both of these relations can be regarded as a basis transformations for the S-matrix or, alternatively, as defining the conformal carrollian theory on null infinity. The idea behind the flat space holography program is to find an independent definition of this conformal carrollian theory.

In the remaining sections of this work, we will use the embedding space formalism to compute two- and three-point functions of conformal carrollian field theories. Based on the above, we can also regard the resulting objects as two- and three-point scattering amplitudes written in an unconventional basis of asymptotic states. For comparison with our results, the explicit expressions for two- and three-point amplitude in the basis \eqref{eq:65} can be found in Appendix \ref{sec:scatt-ampl-moment}.

\section{Scalar Two-Point Correlators}
\label{eq:carcor2pt}

We are now in the position to employ the embedding space formalism of
the last section to the determination of correlators in conformal carrollian
theories. We will start by
deriving the two-point correlator of carrollian boundary scalar fields
\begin{equation}
  \label{eq:twopointdef}
\mathcal{G}^{\Delta_1,\Delta_2}_{2}\equiv   \langle \phi^{(\Delta_1,0)}(u_1,z_1,\zb_1)\phi^{(\Delta_2,0)}(u_2,z_2,\zb_2)\rangle
\end{equation}
from the correlator of embedding space scalar
fields $\Phi^{\Delta_1}(X_1),\Phi^{\Delta_2}(X_2)$. We will assume in the following that they scale
under \eqref{eq:homogscale} with $\Delta_1$ and $\Delta_2$,
respectively, and take the appropriate limits $\Delta_i\to 1$ only at the very end to compare with the expressions coming from \eqref{eq:65} in Appendix \ref{sec:scatt-ampl-moment}.

In the embedding space approach to conventional CFTs, the two point
correlator of scalar fields $\Phi^{\Delta_1}(X_1),\Phi^{\Delta_2}(X_2)$ is
determined by the fact that the only
$\textrm{SO}(4,2)$-invariant quantity available is $X_{12}\equiv X^A_1X^B_2g_{AB}$ and requiring the correct scaling under \eqref{eq:homogscale}, see equation \eqref{eq:12}.
Since we are interested in Poincaré invariant quantities, we have more
structure available in the form of $I^A$. The most general ansatz consistent with the correct scaling and built out of the available quantities $X^A_1,X^A_2,I^A$, and the metric $g_{AB}$ is given by
\begin{equation}
  \label{eq:caransatz}
\langle\Phi^{\Delta_1}(X_1)\Phi^{\Delta_2}(X_2)\rangle=(IX_1)^{-\Delta_1}(IX_2)^{-\Delta_2}f\left(\frac{(IX_1)(IX_2)}{X_{12}}\right)\,,
\end{equation}
where $(IX_i)=I\cdot X_i$.
As expected, Poincaré invariance is less restrictive than full conformal invariance. To proceed, we will assume that $f$ is analytic which is a strong requirement but sufficient to lead to our desired result. We have then
\begin{equation}
  \label{eq:66}
\langle\Phi^{\Delta_1}(X_1)\Phi^{\Delta_2}(X_2)\rangle=Q^{(\Delta_1,\Delta_2)}\equiv\sum^\infty_{n=-\infty}\frac{(IX_1)^{-\Delta_1+n+\delta}(IX_2)^{-\Delta_2+n+\delta}}{X^{n+\delta}_{12}}C_n\Gamma(n+\delta)
\end{equation}
where we made part of the normalization explicit to simplify the
following expressions and we included an arbitrary parameter $\delta$.\footnote{Note that this correlator has the form of a sum over three-point correlators in a conventional CFT where one of the fields is restricted to the point $I^A$, with the sum going over the weights of that particular field.} In our parametrization \eqref{eq:82} this yields
\begin{equation}
  \label{eq:11}
  \langle \Phi^{\Delta_1}(X_1)\Phi^{\Delta_2}(X_2)\rangle
  =\sum^\infty_{n=-\infty}C_n\frac{\Gamma(\delta+n)(\sigma_1)^{-\Delta_1+\delta+n}(\sigma_2)^{-\Delta_2+\delta+n}}{\left[(\sigma_1\sigma_2(r_{12}u_{12}-r_1r_2|z_{12}|^2)\right]^{\delta+n}}\,.
\end{equation}
Following our prescription to the
boundary, i.e., to obtain conformal carrollian fields, we set $r_i=\veps_i \sigma^{-1}_i$.
The correlator for two carrollian fields is then given by
\begin{align}
    \label{eq:14}
 \mathcal{G}^{\Delta_1,\Delta_2}_{2}
  =\sum^\infty_{n=-\infty}\lim_{\sigma_1,\sigma_2 \to 0}
  \frac{C_n\Gamma(\delta+n)\sigma^{-\Delta_1+\delta+n}_1\sigma^{-\Delta_2+\delta+n}_2}{((\veps_1\sigma_2-\veps_2\sigma_1) u_{12}-\veps_{1}\veps_{2}|z_{12}|^2)^{\delta+n}},
\end{align}
where we exchanged the order of sum and limit.\footnote{We note that
  this is of the form of a bulk to boundary propagator in euclidean
  $\textrm{AdS}_3$ if we set $\sigma_1=\sigma_2$, e.g.,  \cite{Witten:1998qj,Kutasov:1999xu}.}

The expression \eqref{eq:66} still needs an appropriate regularization. This can be achieved by rotating $x^0_i\to x^0_i-i \epsilon_i$, where $\epsilon_i$ is Euclidean time. In Euclidean signature, correlators must be time-ordered in Euclidean time. In Lorentzian signature, there is no such requirement on the ordering of operators. Rather, the order is dictated by analytic continuation from time-ordered Euclidean correlators; cf., e.g., \cite{Hartman:2015lfa} for a detailed discussion. The ordering in \eqref{eq:66} therefore requires $\epsilon_1-\epsilon_2\equiv \epsilon_{12}>0$. With this regularization, the denominator in \eqref{eq:14} acquires the additional term
\begin{equation}
  \label{eq:denomreg}
-i \epsilon_{12}(\veps_1\sigma_2 P_1-\veps_2\sigma_1P_2),
\end{equation}
where $P_1=(1+z_1\zb_1)$.

We can already distinguish two cases.
If $|z_{12}|$ is non-zero, the first term in the denominator can be
disregarded. Having a finite correlator picks out the term
$n=\Delta_1-\delta$ and sets $\Delta_1=\Delta_2$ so that one finds
\begin{align}
  \label{eq:uind}
  \mathcal{G}^{(\Delta_1,\Delta_2)}_{2} =\frac{C'\Gamma(\Delta_1)}{\left[-\veps_{1}\veps_{2} |z_{12}|^2\right]^{\Delta_1}}\delta_{\Delta_1,\Delta_2}
\end{align}
This correlator, being of the form of a 2d CFT correlator, is invariant under the
global subgroup of carrollian symmetries \eqref{eq:ccprimary}, even though Poincaré translations act
trivially. We have thus reproduced the correlator of the
$u$-independent branch discussed in
\cite{Chen:2021xkw,Donnay:2022aba,Bagchi:2022emh}. These carrollian correlators appear in the magnetic carrollian limit \cite{Henneaux:2021yzg,Campoleoni:2022ebj,Rivera-Betancour:2022lkc,Gupta:2020dtl}; see also \cite{Baiguera:2022lsw} where it was shown explicitly that a magnetic carrollian theory can be reduced to a lower-dimensional Euclidean CFT.

In order to have a
$u$-dependent correlator we expect to find a correlator proportional to a delta
function in $|z_{12}|^2$. Therefore, the following expressions should all be understood in a distributional sense.\footnote{For a careful analysis of the expression \eqref{eq:14} using the method of regions, see \cite{Donnay:2022ijr}.} One could either take the consecutive limit $\sigma_1\rightarrow 0$, $\sigma_2\rightarrow 0$ (or vice versa) or the simultaneous limit $\sigma_1=\sigma_2=\sigma\to 0$ in expression \eqref{eq:14}. We will consider first the former. Taking $\sigma_1\to 0$, the only term in the sum with a finite contribution comes from $n=\Delta_1-\delta$, that we assume to be an integer, with coefficient $C_{\Delta_1-\delta}\equiv C$. We will truncate the sum to this term. Assuming that $\veps_1=1=-\veps_2$ the coefficient of $-i\epsilon_{12}$ is always positive. We can then use Schwinger parametrization
\begin{equation}
  \label{eq:Schwingerform}
\frac{1}{(A-i\epsilon)^\delta}=\frac{i^\delta}{\Gamma(\delta)}\int^\infty_0 \dd \omega\, \omega^{\delta-1} e^{-i\omega (A-i \epsilon)}
\end{equation}
to write
\begin{equation}
  \label{eq:67}
 \mathcal{G}^{(\Delta_1,\Delta_2)}_{2}
  =\lim_{\sigma_{2} \to 0}C\sigma^{\Delta_1-\Delta_2}_2 i^{\Delta_1}\int^\infty_0\dd \omega \omega^{\Delta_1-1} e^{-i\omega (|z_{12}|^2+ \sigma_2 (u_{12}-i \epsilon_{12}))}\,,
\end{equation}
where we rescaled $\epsilon_{12}\to \epsilon_{12}/(1+|z_1|^2)$.
Rescaling $\omega \rightarrow \omega/\sigma_2$ we can perform a saddlepoint
approximation for small $\sigma_2$ to find
\begin{align}
  \label{eq:3}
  \mathcal{G}^{(\Delta_1,\Delta_2)}_{2}&=2\pi C i^{\Delta_1-1}\!\!\!  \lim_{\sigma_{2}\to 0}\sigma^{1-\Delta_2}_2  \delta(z_{12})\delta(\zb_{12})\,\int^\infty_0\dd \omega\,\omega^{\Delta_1-2} e^{-i\omega(u_{12}-i \epsilon_{12})}\,.
\end{align}
We see that the requirement of having a finite correlator restricts $\Delta_2=1$.
Performing the integral over $\omega$, i.e., undoing the Schwinger parametrization from above we find the expression
\begin{equation}
  \label{eq:halfBagchi} \mathcal{G}^{(\Delta_1,\Delta_2)}_{2}=2\pi C\frac{\Gamma(\Delta_1-1)}{(u_{12}-i\epsilon_{12})^{\Delta_1-1}}\delta(z_{12})\delta(\zb_{12}) \qquad \Delta_2=1\,.
\end{equation}
If we choose to the take the consecutive limit in the other ordering, we would obtain \eqref{eq:halfBagchi} with $\Delta_1$ and $\Delta_2$ switched. It is straightforward to show along the same lines that the simultaneous limit of the expression \eqref{eq:14} leads to
\begin{equation}
  \label{eq:finalBagchi}
 \mathcal{G}^{(\Delta_1,\Delta_2)}_{2}=\pi C\frac{\Gamma(\Delta_1+\Delta_2-2)}{(\veps_{12}u_{12}-i\epsilon_{12})^{\Delta_1+\Delta_2-2}}\delta(z_{12})\delta(\zb_{12})\,,
\end{equation}
with no restrictions on the individual $\Delta$. We thus reproduce, up to an arbitrary normalization, the
time-dependent correlation function branch found in
\cite{Chen:2021xkw,Bagchi:2022emh}.

The correlator \eqref{eq:finalBagchi} depends only on the sum
$\Delta=\Delta_1+\Delta_2$ of the individual scalings. This is also
known from usual CFT correlators with contact terms
\cite{Schwimmer:2018hdl,Nakayama:2019mpz}. The invariance of this correlator can be checked explicitly by acting with the global conformal correlators. It turns out that the
correlator is completely fixed by carrollian transformations and
carrollian dilatations. This also has the surprising consequence that
there is no restriction on individual spins. Instead one finds
$s_{1}+s_{2}=0$, even though one started with a correlator of scalar
fields from the point of view of the embedding spacetime. We thus
expect to find the same result when discussing spinning correlators in
the embedding space in Section \ref{sec:spin2ptfun}.

In order to reproduce the correlator for conformal carrollian fields
dual to fields in asymptotically flat spacetimes according to \cite{Donnay:2022aba,Donnay:2022wvx}, we should specify
\eqref{eq:finalBagchi} to the value $\Delta=\Delta_1+\Delta_2\rightarrow 2$, in which
case the correlator is clearly divergent. Expanding around this value
of $\Delta$ we reproduce the two-point correlator of carrollian boundary fields
discussed in \cite{Liu:2022mne,Donnay:2022wvx}
\begin{align}
  \label{eq:Donnayphys} \mathcal{G}^{(1,1)}_{2}&=\frac{C'}{\Delta-2}\delta(z_{12})\delta(\zb_{12})+C'(\gamma_E+\ln|\veps_{12}u_{12}|-i\pi\Theta(-\veps_{12}u_{12}))\delta(z_{12})\delta(\zb_{12})\,,
\end{align}
again up to an arbitrary constant and a shift in the finite piece proportional to $\delta(z_{12})\delta(\zb_{12})$. The latter can be accounted for by an appropriate choice of the normalization constants in $C_n$.
  As discussed in Appendix \ref{sec:scatt-ampl-moment},
this correlator can be obtained in an appropriate regularization of
the Fourier transform of the norm of a massless two-particle state.
Here we notice that this regularization arises naturally in our
approach. In particular, the integral in \eqref{eq:3} is convergent if
$\Delta_1,\Delta_2>1$, i.e., if the fields obey a slightly stronger
fall-off.\footnote{IR singularities in four-dimensional Minkowski
  space are usually regularized by analytic continuation to $d>4$.
  Note that in higher dimensions one would also find a faster fall-off
  than $\Delta_1=\Delta_2=1$. Thus, the present regularization is
  consistent with dimensional regularization.} For more comments on the interpretation of this object see Appendix \ref{sec:scatt-ampl-moment}.

\section{Spinning Two-Point Correlators}
\label{sec:spin2ptfun}
We saw in the previous section that the correlator
\eqref{eq:finalBagchi} is valid also for spinning particles as long as
$s_1+s_2=0$. We thus expect to find the same result when starting from
the correlator of spinning fields in the embedding space.

Our ansatz for the correlator of a particle with spin one is
\begin{equation}
  \label{eq:spin1ansatz}
  \langle \Phi^{\Delta_1}_{A}(X_1)\Phi^{\Delta_2}_{B}(X_2)\rangle=Q^{(\tau_1,\tau_2)}W_{AB}= \sum_{n}\frac{(I\cdot X_1)^{-\tau_1+\delta+n}(I\cdot X_2)^{-\tau_2+\delta+n}}{(X_1\cdot X_2)^{\delta+n}}W_{AB}\,,
\end{equation}
where $\tau_i=\Delta_i+1$ and $W_{AB}$ scales linearly with $X^A_1,X^A_2$.
Since the fields $\Phi_{A_1}$ obey the conditions \eqref{eq:transverscon}, we have to require
\begin{equation}
  \label{eq:transtensorone}
X^{A}_1W_{AB}=X^{B}_2W_{AB}=I^{A}W_{AB}=I^{B}W_{AB}=0\,.
\end{equation}
If we want to describe the conformal carrollian boundary fields coming from \eqref{eq:Donnaydef} we further need to require
\begin{equation}
  \label{eq:48}
\left(
    \partial^{A}-\frac{(I\cdot \partial)}{(X\cdot I)}X^{A}-(l+2-\Delta_1)I^A
  \right) \langle \Phi^{\Delta_1}_{A}(X_1)\Phi^{\Delta_2}_{B}(X_2)\rangle=0\,,
\end{equation}
with $\Delta_1=\Delta_2=1$,
up to terms that vanish after projecting to the physical field, and similar on the second index. This will, in general, impose an additional condition on $W_{AB}$ that we will return to later.

To construct $W_{AB}$ we have at our disposal the quantities $g_{AB},X^A_{1},X^A_2,I^A$. Starting from an arbitrary tensor $\tilde{W}_{AB}$ we can achieve the conditions \eqref{eq:transtensorone} by projecting with
\begin{equation}
  \label{eq:deftransproj}
  \Pi_{AB}(X)=g_{AB}-\frac{I_A X_{B}+I_B X_{A}}{I\cdot X},\qquad I^A\Pi_{AB}=0,\qquad X^A\Pi_{AB}=0\,.
\end{equation}
The most general ansatz for $W_{AB}$ is then given by
\begin{equation}
  \label{eq:17}
  W_{AB}=\Pi(X_1)\indices{_A^C}\tilde{W}_{CD}\Pi(X_2)\indices{_B^D}\qquad \tilde{W}_{AB}=g_{AB} X_{12}+\alpha_1X_{2A}X_{1B}\,,
\end{equation}
where $\alpha_1$ is, for the moment, arbitrary and we included only terms in $\tilde{W}$ that do not vanish under projection with $\Pi$. The tensor constructed in this way will satisfy \eqref{eq:transtensorone} in accordance with \eqref{eq:transverscon}. Finally, note that in the projection to the physical tensor \eqref{eq:centralchain} we have
\begin{equation}
  \label{eq:52}
  \partial_zX^A_1I_A=\partial_zX^B_2I_B=\partial_zX^A_1X_{1A}=\partial_zX^B_2X_{1B}=0\,,
\end{equation}
so that we can drop all of these terms from $W$ for the evaluation of the physical correlator. Since the terms generated by $\Pi$ from $\tilde{W}$ are exactly of this form, the difference between $\tilde{W}$ and $W$ is irrelevant for the evaluation of the physical correlator.

Projecting now to the physical fields we have, using our prescription,
\begin{equation}
  \label{eq:23}
\partial_{z_1}X^A_1\partial_{\zb_2}X^B_2g_{AB}\rightarrow \veps_1\veps_2\qquad \partial_{z_1}X^A_1\partial_{\zb_2}X^B_2X_{2A}X_{1B}\rightarrow -|z_{12}|^2\,.
\end{equation}
We already see that the first term will yield the same result as in the scalar case. For the second term we can use
\begin{equation}
  \label{eq:trickyone}
\frac{-|z_{12}|^2}{(X_1\cdot X_2)^{(\delta+n)}}=-\frac{1}{(\delta+n-1) r_1r_2 \sigma_1\sigma_2}z_{12}\partial_{z_{12}}\frac{1}{(X_1\cdot X_2)^{\delta+n-1}}\,.
\end{equation}
Using our prescription for obtaining the carrollian correlator, we find that this term, understood in a distributional sense, yields the same contribution as in the scalar case with coefficient $\veps_1\veps_2\alpha_1/(\delta+n-1)$. Mirroring the steps for the scalar correlator, the only finite term in the sum is $n=\tau_1+\tau_2-1-\delta=\Delta+1-\delta$. Taken together, the result for the correlator in the simultaneous limit is
\begin{equation}
  \label{eq:Bagchicorspin}
\langle \phi^{(\Delta_1,1)}_{z}\phi^{(\Delta_2,1)}_{\zb}\rangle=-C'(1+\frac{\alpha_1}{\Delta})\frac{\Gamma(\Delta-2)}{(\veps_{12}u_{12}-i\epsilon)^{\Delta-2}}\delta(z_{12})\delta(\zb_{12})\,,
\end{equation}
with corresponding results in the consecutive limits.
Thus, starting with the two-point function for spinning particles in embedding space leads, up to the coefficient, to the same correlator \eqref{eq:finalBagchi} with the same restriction $s_1+s_2=0$.

Let us now examine the condition \eqref{eq:48}. We have
\begin{equation}
  \label{eq:59}
  \left(
    \frac{\partial}{\partial X_{1A}}-\frac{(I\cdot \partial)}{(X_1\cdot I)}X^{A}_1\right)(Q^{\tau_1,\tau_2}W_{AB})=O(X_{2B},I_{B})\,,
\end{equation}
by which we mean the right hand side is zero up to terms vanishing after the projection. Note that we dropped the term proportional to $I^A$ due to the condition \eqref{eq:transtensorone}. Replacing $W$ by $\Pi \tilde{W} (\Pi)^T$ one finds
\begin{equation}
  \label{eq:60}
\left(
    \frac{\partial}{\partial X_{1A}}-\frac{(I\cdot \partial)}{(X_1\cdot I)}X^{A}_1-\frac{1}{(I\cdot X_1)}(3-\Delta_1)I^A\right)(Q^{\tau_1,\tau_2}\tilde{W}_{AB})=O(X_{2B},I_{B})\,.
\end{equation}
Explicitly evaluating the derivative on our ansatz \eqref{eq:17} we find the relation $\alpha_1=\Delta/(5-2\Delta)$, which fixes the form of the correlator \eqref{eq:spin1ansatz} up to an overall normalization. In particular, in the case of $\Delta=\Delta_1+\Delta_2\rightarrow 2$ we have $\alpha_1=2$.

We do not expect correlators of higher-valued tensor fields in embedding space to yield correlators other than \eqref{eq:Bagchicorspin} up to a coefficient. This was shown explicitly in \cite{Bagchi:2022emh,Donnay:2022wvx} by checking the conformal carrollian Ward identities. In order to show this in the embedding space, a more economic approach for the construction of the tensors, such as the one in \cite{Costa:2011mg}, and the implementation of the condition \eqref{eq:48} would be desireable. We will have more comments on this in the discussion.

\section{Scalar Three-Point Corrrelators}
\label{sec:scalar-fields}

A generic three-point function for carrollian fields vanishes in Lorentzian
signature. This is for the same reason that the
three-point amplitude vanishes in Lorentzian signature due to special
three-particle kinematics.
It is well-known that this can be
circumvented by either complexifying the momenta of external particles
or by going to split signature. Although zero in Lorentzian signature, the three-point correlators obtained in this way serve as fundamental building blocks for higher-point, non-vanishing amplitudes; see e.g. \cite{Elvang:2015rqa}.
We will follow a similar strategy here.

 The most general
ansatz consistent with Poincaré invariance and scaling invariance in our embedding space approach is
proportional to an arbitary function of the three invariants
$(I\cdot X_i)(I\cdot X_j)X^{-1}_{ij}$ times appropriate powers of
$(I\cdot X_i)$ that fix the scaling. As in the case of the two-point function, we assume that this function can be expanded in a power series so that one finds
\begin{align}
  \label{eq:threepartans}
  \langle &\Phi^{\Delta_1}_{1}(X_1)\Phi^{\Delta_2}_{2}(X_2)\Phi^{\Delta_3}_{3}(X_3)\rangle
  = Q^{(\Delta_1,\Delta_2,\Delta_3)}\\
  &\equiv \sum_{\delta_1,\delta_2,\delta_3}C_{\delta_1\delta_2\delta_3}\frac{\Gamma(\delta_1)\Gamma(\delta_2)\Gamma(\delta_3)}{(X_{12})^{\delta_3}(X_{23})^{\delta_1}(X_{13})^{\delta_2}}
  (I X_1)^{-\Delta_1+\delta_{23}}(I X_2)^{-\Delta_2+\delta_{13}}(I X_3)^{-\Delta_3+\delta_{12}}\,,\nonumber
\end{align}
where $\delta_{ij}=\delta_i+\delta_j$,
$\delta_{123}=\delta_1+\delta_2+\delta_3$ and we made part of the coefficient function $C_{\delta_1\delta_2\delta_3}$ explicit to simplify the following formulas. The summations, that we wrote for brevity as $\sum_{\delta_i}$, should be understood as in \eqref{eq:11}, i.e., as sums over $n_i$ after the shift $\delta_i\rightarrow \delta_i+n_i$.

Using the parametrization \eqref{eq:82} and our prescription \eqref{eq:centralchain} we want to obtain from this the conformal carrollian correlator
\begin{equation}
  \label{eq:threepointabbrev}
\mathcal{G}^{(\Delta_1,\Delta_2,\Delta_3)}_{3}\equiv\langle
\mathcal{P}^{(\veps_1,\veps_2,\veps_3)}[\phi^{(\Delta_1,0)}(u_1,z_1,\zb_1)\phi^{(\Delta_2,0)}(u_2,z_2,\zb_2)\phi^{(\Delta_3,0)}(u_3,z_3,\zb_3)]\rangle,
\end{equation}
where we assumed scaling weights $\Delta_i$ for the fields. The notation $\mathcal{P}^{(\veps_1,\veps_2,\veps_3)}$ denotes that the operators have a particular order depending on $\veps_i$ that we will make more precise below.

As in the case of the two point function, one can distinguish several cases.\footnote{This expression, understood as a distribution, can again be analyzed using the method of regions \cite{Smirnov:2002pj}.} If all $z_{ij}\zb_{ij}$ are finite, then the limit $\sigma_i \to 0$ can be taken freely, and \eqref{eq:threepartans} reduces to a $u$-independent correlator. The requirement of having a finite correlator fixes $\delta_i=\frac{1}{2}(\Delta_i-\Delta_j-\Delta_k)$ so that one finds the usual three-point correlator of a CFT. As mentioned above, this branch corresponds to magnetic carrollian theories.

Let us now turn to genuine conformal carrollian correlators that transform under the full Poincaré group and not just its Lorentz subgroup. In order not to clutter the expressions, we will consider a single term in the sum \eqref{eq:threepartans} and assume that all limits can be freely commuted through the summations. In Schwinger parametrizations we have then again
 \begin{align}
   \label{eq:26}
   \mathcal{G}^{(\Delta_1,\Delta_2,\Delta_3)}_{3}&= C i^{\delta_{123}}\lim_{\sigma_{1,2,3} \to 0}\prod_{\circlearrowleft_{ijk}}\sigma^{-\Delta_i+\delta_j+\delta_k}_i
   \int^\infty_0\dd \omega_1\dd \omega_2\dd \omega_3\omega^{\delta_1-1}_1\omega^{\delta_2-1}_2\omega^{\delta_3-1}_3 \nonumber \\
   &\qquad\exp\left(
     -i\sum_{\circlearrowleft_{ijk}}\omega_i\left(\veps_j\veps_k|z_{jk}|^2\right)
     -i\sum_{\circlearrowleft_{ijk}}\omega_i\left(\veps_j\sigma_k-\veps_k\sigma_j\right)u_{jk}\right)\\
   &\qquad\exp\left(-\sum_{\circlearrowleft_{ijk}}\omega_i\epsilon_{jk}\left(\veps_j\sigma_k P_j-\veps_k\sigma_jP_k\right)\right)\nonumber\,,
 \end{align}
 where the term in the last line comes from our $i\epsilon$ prescription \eqref{eq:denomreg}.

 As before, let us consider the consecutive limits $\sigma_3\to 0,\sigma_2\to 0,\sigma_1\to 0$. The requirement of having a convergent integral in \eqref{eq:26} fixes a certain ordering of the $\epsilon_i$, and thus the operators on the left hand side, depending on the $\veps_i$. Requiring the coefficient of the last exponential in \eqref{eq:26} to have a positive sign in the chosen consecutive limits leads to restrictions on the Euclidean times $\epsilon_i$ and therefore to the following operator ordering in \eqref{eq:threepointabbrev}:
 \begin{center}
 \begin{tabular}{ccc|c}
   $\veps_1$ & $\veps_2$ &$\veps_3$& $\mathcal{P}[\Phi_1\Phi_2\Phi_3]$\\
   \hline
   $+1$&$+1$&$-1$& $\Phi_2\Phi_1\Phi_3$\\
   $+1$&$-1$&$+1$& $\Phi_3\Phi_1\Phi_2$\\
   $+1$&$-1$&$-1$& $\Phi_1\Phi_2\Phi_3$,
 \end{tabular}
 \end{center}
 and the inverse ordering if the signs of all $\veps_i$ are flipped. Note that operators with $\veps_i=+1(-1)$, i.e., operators defined at future (past)null infinity, always stand to the left (right).

As for the two-point function, the requirement of having a finite correlator fixes $\delta_2=-\delta_1+\Delta_3,\delta_3=-\delta_1+\Delta_2$ so that we can set all other coefficients in \eqref{eq:threepartans} to zero. After a rescaling $\omega_i\rightarrow \omega_i/\sigma_1$ one finds
 \begin{align}
   \label{eq:26a}
   \mathcal{G}^{(\Delta_1,\Delta_2,\Delta_3)}_{3}&= C i^{\delta_{123}}\lim_{\sigma_{1} \to 0}\sigma^{-\Delta_1-\delta_1}_1
   \int^\infty_0\dd \omega_1\dd \omega_2\dd \omega_3\omega^{\delta_1-1}_1\omega^{\delta_2-1}_2\omega^{\delta_3-1}_3 \nonumber \\
   &\qquad\exp\left(
     -i\sum_{\circlearrowleft_{ijk}}\frac{\omega_i}{\sigma_1}\left(\veps_j\veps_k|z_{jk}|^2\right)
     -i\left(-\omega_3\veps_2u_{12}+\omega_2\veps_3u_{31}\right)\right)\,,
 \end{align}
 where we dropped the $-i\epsilon$ prescription.
 Evaluating the
 integral in a saddlepoint approximation one can solve for the saddle
 point equations by assuming that either all $z_{ij}\neq 0$ or
 all $\zb_{ij}\neq 0$. This is clearly not possible in Lorentzian signature where $z$ and $\zb$ are complex conjugates of each other. We will therefore analytically continue the coordinates on null infinity by considering $z$ and $\zb$ real and independent. This corresponds to switching over to
a pseudo-conformal carrollian structure, where the degenerate spatial metric has signature $(1,1)$.
 Choosing  $z_{ij}\neq 0$ one finds
\begin{align}
  \label{eq:28}
   \mathcal{G}^{(0)}_3&=\lim_{\sigma_3 \to 0}\frac{(2\pi)^2C i^{\delta_{123}-2}\sigma^{2-\Delta_1-\delta_1}}{|z_{12}||z_{31}|}
   \left(\frac{\epsilon_3z_{23}}{\epsilon_1z_{12}}\right)^{\delta_3-1}\left(\frac{\epsilon_2z_{23}}{\epsilon_1z_{31}}\right)^{\delta_2-1} \delta(\zb_{13})\delta(\zb_{23}) \nonumber\\
&   \int^\infty_0\dd \omega \omega^{\delta_{123}-3}
                         \exp\left(-i\frac{\omega z_{23}\veps_1\veps_2\veps_3}{z_{12}z_{31}}(u_1z_{23}+u_2z_{31}+u_3z_{12})\right) \Theta\left(\frac{\epsilon_2z_{23}}{\epsilon_1z_{31}}\right) \Theta\left(\frac{\epsilon_3z_{23}}{\epsilon_1z_{12}}\right)\,.
\end{align}
Here, we solved the saddle-point equations in terms of $\omega_2,\omega_3,\zb_2,\zb_3$. The $\Theta$ functions in the above expression are necessary since $\omega_2,\omega_3$ are required to be greater than zero. Note that here and in the following $|z_{ij}|$ denotes the absolute value of $z_{ij}$ which is assumed to be real. Finally, we obtain the restriction $\delta_1=2-\Delta_1$ that eliminates the last sum in \eqref{eq:threepartans}. Rescaling $\omega \rightarrow \omega z_{12}z_{31}/(\veps_2\veps_3z_{23}^2)$, which is always positive due to the $\Theta$ functions, and performing the integral that is of the same form as in \eqref{eq:3} we finally find
\begin{align}
  \label{eq:final3ptscalar}
  \mathcal{G}^{(\Delta_1,\Delta_2,\Delta_3)}_3&=\frac{(2\pi)^2C}{|z_{12}||z_{31}|}
\left(\frac{\epsilon_3z_{23}}{\epsilon_1z_{12}}\right)^{1-\delta_{12}}\left(\frac{\epsilon_2z_{23}}{\epsilon_1z_{31}}\right)^{1-\delta_{31}} \delta(\zb_{13})\delta(\zb_{23})\Theta\left(\frac{\epsilon_2z_{23}}{\epsilon_1z_{31}}\right) \Theta\left(\frac{\epsilon_3z_{23}}{\epsilon_1z_{12}}\right)\,\nonumber\\
                                              &\qquad \frac{\Gamma(\delta_{123}-2)}{\left(\frac{\veps_1}{z_{23}}(u_1z_{23}+u_2z_{31}+u_3z_{12})\right)^{\delta_{123}-2}}\\
  &=\frac{(2\pi)^2C}{|z_{12}||z_{31}|}
\left(\frac{\epsilon_3z_{23}}{\epsilon_1z_{12}}\right)^{1-\Delta_3}\left(\frac{\epsilon_2z_{23}}{\epsilon_1z_{31}}\right)^{1-\Delta_2} \delta(\zb_{13})\delta(\zb_{23})\Theta\left(\frac{\epsilon_2z_{23}}{\epsilon_1z_{31}}\right) \Theta\left(\frac{\epsilon_3z_{23}}{\epsilon_1z_{12}}\right)\,\nonumber\\
                                              &\qquad \frac{\Gamma(\Delta-4)}{\left(\frac{\veps_1}{z_{23}}(u_1z_{23}+u_2z_{31}+u_3z_{12})\right)^{\Delta-4}},.
\end{align}
It can be checked explicitly that this correlator obeys the global Ward identities of a conformal carrollian correlator of fields with weight $\Delta_i$ and spin zero. In the limit $\Delta_{1,2,3}\to 1$ the correlator is again divergent due to the pole in the Gamma function, but we recognize from
\begin{align}
  \label{eq:final3ptbound}
  \langle \mathcal{P}^{(\veps_1,\veps_2,\veps_3)}[\phib^{(0)} \phib^{(0)} \phib^{(0)}]\rangle&=\frac{1}{3-\Delta}\frac{(2\pi)^2C}{|z_{12}||z_{31}|}
  \delta(\zb_{13})\delta(\zb_{23})\Theta\left(\frac{\epsilon_2z_{23}}{\epsilon_1z_{31}}\right) \Theta\left(\frac{\epsilon_3z_{23}}{\epsilon_1z_{12}}\right)\\
  &\qquad \left(\frac{\veps_1}{z_{23}}(u_1z_{23}+u_2z_{31}+u_3z_{12})\right)+O(\Delta^0)\,,
                       \end{align}
                       the 3 point scattering amplitude for three scalar fields from \eqref{eq:43} proportional to a delta function.

\section{Spinning Three-Point Correlators}
\label{sec:3ptfun}
We will take as our ansatz
\begin{align}
  \label{eq:spinning ansatz}
  \langle \Phi^{\Delta_1}_{A_1\ldots A_{s_{1}}}(X_1)\Phi^{\Delta_2}_{B_1\ldots B_{s_2}}(X_2)\Phi^{\Delta_3}_{C_1\ldots C_{s_3}}(X_3)\rangle=\sum_k Q^{(\tau_1,\tau_2,\tau_3)}_kW^{(s_1,s_2,s_3),(k)}_{A_1\ldots A_{s_1}B_1\ldots B_{s_2}C_1\ldots C_{s_3}}
\end{align}
with the $Q^{(\tau_1,\tau_2,\tau_3)}_k$ of the same form as in \eqref{eq:threepartans} but with the replacements $\Delta_i\rightarrow\tau_i=\Delta_i+s_i$ and $W^{(s_1,s_2,s_3)}$ is of weight $s_i$ in $X_i$. Note that we have a sum over different tensor structures $W$ as in the case of the two-point function, but the coefficients can be arbitrary functions of the form of the three-point scalar function.\footnote{As before, we can think of this correlator as a CFT 4-point function of fields at the points $X_{1},X_2,X_3,I$ where we sum over the weight of the field at $I$ and series-expand the undetermined function of conformal cross-ratios.}

We will restrict ourselves to the simplest three-point correlators with spin.
\subsection{Spin 1, scalar, scalar}
We will start with the simplest case of one spinning field in embedding space, so that we have $s_1=1$, $s_2=s_3=0$.

As in the case of the two-point function, we will construct the most general tensor $\tilde{W}$ out of the available quantities $I,X_1X_2,X_3$ and later use the projection operator \eqref{eq:deftransproj} to construct the tensor $W=\Pi \tilde{W}$ such that the correlator obeys the desired conditions $I^AW_A=X^{A}_1W_A=0$.
The most general tensor consistent with the required scaling implied by the ansatz \eqref{eq:spinning ansatz} is
\begin{align}
  \label{eq:24}
  \langle \Phi^{\Delta_1}_{A}(X_1)\Phi^{\Delta_2}(X_2)&\Phi^{\Delta_3}(X_3)\rangle\\
  &=\Pi\indices{^B_A}(X_1)\left(Q^{(\Delta_1+1,\Delta_2,\Delta_3)}_1X_{2B}\frac{(I X_1)}{(I X_2)}+Q^{(\Delta_1+1,\Delta_2,\Delta_3)}_2 X_{3B}\frac{(IX_1)}{(IX_3)}\right)\nonumber
\end{align}
where the $Q^{(\Delta_1+1,\Delta_2,\Delta_3)}_k$ are of the same form as \eqref{eq:threepartans} but with different expansion coefficients $C_{\delta_1\delta_2\delta_3}$. As discussed in Section \ref{sec:spin2ptfun}, we need the projector so that the correlator obeys the condition \eqref{eq:transverscon}, but the additional terms will not be relevant when projecting to the physical fields.

The correlator can be straightforwardly evaluated referring to the discussion in the previous section. The additional factors in \eqref{eq:24} can be accommodated for by a shift in the scaling weights so one can equivalently write
\begin{equation}
  \label{eq:5}
    \langle \Phi^{\Delta_1}_{A}(X_1)\Phi^{\Delta_2}(X_2)\Phi^{\Delta_3}(X_3)\rangle=\Pi\indices{^B_A}(X_1)\left(Q^{(\Delta_1,\Delta_2+1,\Delta_3)}_1X_{2B}+Q^{(\Delta_1,\Delta_2,\Delta_3+1)}_2 X_{3B}\right)\,.
  \end{equation}
  For the same reason as in the last section the sums in $Q$ will reduced to a single term if one is interested in finite correlators.
  To find the correlator of carrollian fields we have to choose to project either with $\partial_{\zb_1}X^A_1$ or $\partial_{z_1}X^A_1$, corresponding to a correlator of fields of helicity $-1,0,0$ or $+1,0,0$. With the choice of resolving the delta functions on the barred sector in \eqref{eq:final3ptscalar}, we will have to take the former option in order to find a finite result.

Using the inner products
\begin{equation}
  \label{eq:22}
  \partial_{\zb_1}X^A_1X_{A2}=-\veps_1\veps_2z_{12}\qquad \partial_{\zb_1}X^A_1X_{A3}=+\veps_1\veps_3z_{31}\,,
\end{equation}
and the result \eqref{eq:final3ptscalar} we have then
\begin{align}
  \label{eq:final100}
\langle  \mathcal{P}^{(\veps_1,\veps_2,\veps_3)}&[\phi^{(\Delta_1,1)}_{\zb}\phi^{(\Delta_2,0)}\phi^{(\Delta_3,0)}]\rangle=(C_2-C_1)\frac{(2\pi)^2\veps_2\veps_3z_{23}}{|z_{12}||z_{31}|}
  \left(\frac{\veps_3z_{23}}{\veps_1z_{12}}\right)^{-\Delta_3}\left(\frac{\veps_2z_{23}}{\veps_1z_{31}}\right)^{-\Delta_2} \\ &\delta(\zb_{13})\delta(\zb_{23})\frac{\Gamma(\Delta-3)}{\left(\frac{\veps_1}{z_{23}}(u_1z_{23}+u_2z_{31}+u_3z_{12})\right)^{\Delta-3}}\Theta\left(\frac{\epsilon_2z_{23}}{\epsilon_1z_{31}}\right) \Theta\left(\frac{\epsilon_3z_{23}}{\epsilon_1z_{12}}\right)\nonumber\,,
\end{align}
with $C_k$ being the two coefficients in the expansions for $Q^{(\tau_1,\tau_2,\tau_3)}_k$ being picked out by the requirement of having a finite correlator. Specifying to the case $\Delta_i\to 1$ this becomes
\begin{align}
  \label{eq:27}
 \langle \mathcal{P}^{(\veps_1,\veps_2,\veps_3)}[\bar{\phi}^{(1)}_{\zb}\bar{\phi}^{(0)}\bar{\phi}^{(0)}]\rangle&=(C_2-C_1)\frac{\sgn(z_{12}z_{23}z_{31})}{|z_{23}|} \delta(\zb_{13})\delta(\zb_{23})\Theta\left(\frac{\epsilon_2z_{23}}{\epsilon_1z_{31}}\right) \Theta\left(\frac{\epsilon_3z_{23}}{\epsilon_1z_{12}}\right)\\
 &\lim_{\Delta\to 3}\frac{\Gamma(\Delta-3)}{\left(\frac{\veps_1}{z_{23}}(u_1z_{23}+u_2z_{31}+u_3z_{12})\right)^{\Delta-3}}\nonumber\,,
\end{align}
where we made use of the Theta functions to rewrite $\veps_2\veps_3\sgn z_{23}=\sgn(z_{12}z_{23}z_{31})$. This is indeed the well-known form of a scattering amplitude of a spin one particle and two scalar fields as one finds e.g. in scalar QED, written in asymptotic position space; see Appendix \ref{sec:scatt-ampl-moment}.

We have not yet imposed the De Donder gauge condition \eqref{eq:DeDondersix}. In the consecutive limit, the De Donder gauge condition yields the equation
\begin{equation}
  \label{eq:89}
  C_1(\Delta_2+\Delta_1-2) X_{13}+C_2(\Delta_3+\Delta_1-2) X_{12}=0\,,
\end{equation}
which is trivially satisfied for $\Delta_i\to 1$.
\subsection{Spin 1, Spin 1, scalar}
\label{sec:spin110}
In this case, the most general tensor is of the form
\begin{equation}
  \label{eq:42}
  \langle \Phi^{\Delta_1}_A(X_1)\Phi^{\Delta_2}_B(X_2)\Phi^{\Delta_3}(X_3)\rangle=\sum^5_{k=1}Q^{(\Delta_1+1,\Delta_2+1,\Delta_3)}_k\tilde{W}^{k}_{CD}\Pi\indices{^C_A}(X_1)\Pi\indices{^C_A}(X_2)\,,
\end{equation}
where the tensors $W^{k}_{AB}$ are given by
\begin{align}
  \label{eq:51}
  \tilde{W}^{(1)}=X_{2A}X_{1B}\,,&\qquad \tilde{W}^{(2)}=X_{2A}X_{3B}\frac{(I X_1)}{(I X_3)}\,,\qquad \tilde{W}^{(3)}=X_{3A}X_{1B}\frac{(IX_2)}{(I X_3)}\,\\
  & \tilde{W}^{(4)}=X_{3A}X_{3B}\frac{(I X_1)^2(I X_2)^2}{(I X_3)^2}\qquad \tilde{W}^{(5)}=X_{12}g_{AB}\,.\nonumber
\end{align}
As above, we can rewrite expression \eqref{eq:42} as
\begin{align}
  \label{eq:53} \langle &\Phi^{\Delta_1}_C(X_1)\Phi^{\Delta_2}_D(X_2)\Phi^{\Delta_3}(X_3)\rangle=\Pi\indices{^A_C}\Pi\indices{^B_D}\Big(Q^{(\Delta_1+1,\Delta_2+1,\Delta_3)}_1X_{2A}X_{1B}+Q^{(\Delta_1,\Delta_2,\Delta_3)}_5 g_{AB}\\ & +Q^{(\Delta_1,\Delta_2+1,\Delta_3+1)}_2X_{2A}X_{3B}
  +Q^{(\Delta_1+1,\Delta_2,\Delta_3+1)}_3X_{3A}X_{1B}+Q^{(\Delta_1,\Delta_2,\Delta_3+2)}_4X_{3A}X_{3B}\Big)\nonumber
\end{align}
In order to compute the correlator between two fields of helicity $-1$ and a scalar field, we project with $\partial_{\zb_1}X^A_1,\partial_{\zb_2}X^A_2$ and find after the consecutive limit
\begin{align}
  \label{eq:final3ptspin110}
  \langle \mathcal{P}^{(\veps_1,\veps_2,\veps_3)}[\phi^{(\Delta_1,1)}_{\zb}\phi^{(\Delta_2,1)}_{\zb}\phi^{(\Delta_3,0)}]\rangle&=\frac{(2\pi)^2(-C_1+C_2+C_3-C_4)z^2_{12}}{|z_{12}||z_{31}|}
            \left(\frac{\epsilon_3z_{23}}{\epsilon_1z_{12}}\right)^{1-\Delta_3}\\ \left(\frac{\epsilon_2z_{23}}{\epsilon_1z_{31}}\right)^{-\Delta_2}\delta(\zb_{13})\delta(\zb_{23})\nonumber
  &\Theta\left(\frac{\epsilon_2z_{23}}{\epsilon_1z_{31}}\right) \Theta\left(\frac{\epsilon_3z_{23}}{\epsilon_1z_{12}}\right)\,
   \frac{\Gamma(\Delta-2)}{\left(\frac{\veps_1}{z_{23}}(u_1z_{23}+u_2z_{31}+u_3z_{12})\right)^{\Delta-2}}\,.
\end{align}
Setting $\Delta_i=1$, this reproduces the Fourier transform of the momentum space correlator for two particles of helicity zero with a scalar particle; cf. Appendix \ref{sec:scatt-ampl-moment}
\begin{align}
  \label{eq:final3ptspin110bdy}
  \langle \mathcal{P}^{(\veps_1,\veps_2,\veps_3)}[\bar{\phi}^{(1)}_{\zb}\bar{\phi}^{(1)}_{\zb}\bar{\phi}^{(0)}]\rangle&=(2\pi)^2(-C_1+C_2+C_3-C_4)\frac{|z_{12}|}{|z_{23}|}\delta(\zb_{13})\delta(\zb_{23})\nonumber
  \\&\Theta\left(\frac{\epsilon_2z_{23}}{\epsilon_1z_{31}}\right) \Theta\left(\frac{\epsilon_3z_{23}}{\epsilon_1z_{12}}\right)\,
   \frac{1}{\left(\frac{\veps_1}{z_{23}}(u_1z_{23}+u_2z_{31}+u_3z_{12})\right)}\,,
\end{align}
where we used the $\Theta$ function to write $\veps_1\veps_2\sgn(z_{31}z_{23})=1$.
This amplitude is produced by a Higgs-gluon fusion operator in the massless limit \cite{Berger:2006sh}.

One could also project \eqref{eq:53} using $\partial_{\zb_1}X^A_1,\partial_{z_2}X^A_2$ to obtain the correlator corresponding to the scattering of two fields of helicity minus one and plus one off a scalar field. We obtain
\begin{align}
  \label{eq:61}
  \langle \phi^{(\Delta_1,1)}_{\zb}\phi^{(\Delta_2,1)}_{z}\phi^{(\Delta_3,0)}\rangle&=\frac{(2\pi)^2C_5\veps_1\veps_2}{|z_{12}||z_{31}|}
            \left(\frac{\epsilon_3z_{23}}{\epsilon_1z_{12}}\right)^{1-\Delta_3}\left(\frac{\epsilon_2z_{23}}{\epsilon_1z_{31}}\right)^{1-\Delta_2} \delta(\zb_{13})\delta(\zb_{23})\nonumber
  \\&\Theta\left(\frac{\epsilon_2z_{23}}{\epsilon_1z_{31}}\right) \Theta\left(\frac{\epsilon_3z_{23}}{\epsilon_1z_{12}}\right)\,
   \frac{\Gamma(\Delta-4)}{\left(\frac{\veps_1}{z_{23}}(u_1z_{23}+u_2z_{31}+u_3z_{12})\right)^{\Delta-4}}\,.
\end{align}
This scattering amplitude has total helicity $h_1+h_2+h_3=0$ and is thus not covered by the little group analysis that leads to \eqref{eq:spinhel3pt}. It can be shown that one cannot construct consistent four-point amplitudes using these interactions. The associated coupling must therefore vanish \cite{McGady:2013sga,Arkani-Hamed:2017jhn}.

Imposing the De Donder gauge condition \eqref{eq:DeDondersix} on the ansatz \eqref{eq:42} for the boundary carrollian fields $\bar{\phi}^{(s)}$ with $\Delta_i=1$ leads to the following conditions in the consecutive limit:
\begin{equation}
  \label{eq:90}
(-2+\Delta_1+\Delta_3)\alpha_5=(-2+\Delta_2+\Delta_3)\alpha_5=0\,, \Delta_1,\Delta_2\to 1
\end{equation}
which are, again, trivially satisfied.
\subsection{Spin 1, Spin 1, Spin 1}

The last case we want to discuss is the three-point correlator of three massless particles of spin one. In order to simplify the discussion, we will focus on tensor structures that are totally antisymmetric under the exchange of any two points. The ansatz for the correlator is then
\begin{equation}
  \label{eq:spin111ansatz}
  \langle \Phi^{\Delta_1}_A(X_1)\Phi^{\Delta_2}_B(X_2)\Phi^{\Delta_3}_C(X_3)\rangle=\sum^{3}_{k=1}Q^{(\Delta_1+1,\Delta_2+1,\Delta_3+1)}_k\tilde{W}^{(k)}_{DEF}\Pi\indices{^D_A}(X_1)\Pi\indices{^E_B}(X_2)\Pi\indices{^F_C}(X_2)\,
\end{equation}
with the tensors
\begin{subequations}
\begin{align}
  \label{eq:spin111tensors}
  \tilde{W}^{(1)}_{ABC}&=X_{2A}X_{3B}X_{1C}+X_{3A}X_{1B}X_{2C}\\
  \tilde{W}^{(2)}_{ABC}&=X_{1B}X_{1C}\frac{X_{2A}(IX_3)-X_{3A}(IX_2)}{(IX_1)}-X_{2A}X_{2C} \frac{X_{1B}(IX_3)-X_{3B}(IX_1)}{(IX_2)}\nonumber\\
  &+X_{3A}X_{3B} \frac{X_{1C} (IX_2)-X_{2C}(IX_1)}{(IX_3)}\\
  \tilde{W}^{(3)}_{ABC}&=g_{AB}(IX_3)((IX_1)X_{2C}-X_{1C}(IX_2))+g_{BC}(IX_1)((IX_2)X_{3A}-(IX_3)X_{2A}) \nonumber\\
  &+g_{AC}(IX_2)(X_{1B}(IX_3)-X_{3B}(IX_1))
\end{align}
\end{subequations}

We will consider first the projection by $\partial_{\zb}X^A_1,\partial_{\zb}X^B_2,\partial_{\zb}X^C_3$ to obtain the three-point correlator of three fields of helicity minus one. Going through the same steps as above we obtain the result
\begin{align}
  \label{eq:63}
\langle \mathcal{P}^{(\veps_1,\veps_2,\veps_3)}[\phi^{(\Delta_1,1)}_{\zb}\phi^{(\Delta_2,1)}_{\zb}\phi^{(\Delta_3,1)}_{\zb}]\rangle&=\frac{(2\pi)^26C_2z_{12}z_{23}z_{31}}{|z_{12}||z_{31}|}
            \left(\frac{\epsilon_3z_{23}}{\epsilon_1z_{12}}\right)^{-\Delta_3}\left(\frac{\epsilon_2z_{23}}{\epsilon_1z_{31}}\right)^{-\Delta_2} \delta(\zb_{13})\delta(\zb_{23})\nonumber
  \\&\Theta\left(\frac{\epsilon_2z_{23}}{\epsilon_1z_{31}}\right) \Theta\left(\frac{\epsilon_3z_{23}}{\epsilon_1z_{12}}\right)\,
   \frac{\Gamma(\Delta-1)}{\left(\frac{\veps_1}{z_{23}}(u_1z_{23}+u_2z_{31}+u_3z_{12})\right)^{\Delta-1}}\,.
\end{align}
In the limit $\Delta_i\to 1$ we reproduce the correct functional form of the three-point amplitude
\begin{align}
  \label{eq:spin111bdyallminus}
\langle \mathcal{P}^{(\veps_1,\veps_2,\veps_3)}[\bar{\phi}^{(1)}_{\zb}\bar{\phi}^{(1)}_{\zb}\bar{\phi}^{(1)}_{\zb}]\rangle&=(2\pi)^2\sgn(z_{12}z_{23}z_{31})6C^{(2)}\frac{|z_{12}z_{31}|}{|z_{23}|}
             \delta(\zb_{13})\delta(\zb_{23})\nonumber
  \\&\Theta\left(\frac{\epsilon_2z_{23}}{\epsilon_1z_{31}}\right) \Theta\left(\frac{\epsilon_3z_{23}}{\epsilon_1z_{12}}\right)\,
   \lim_{\Delta\to 3}\frac{\Gamma(\Delta-1)}{\left(\frac{\veps_1}{z_{23}}(u_1z_{23}+u_2z_{31}+u_3z_{12})\right)^{\Delta-1}}\,,
\end{align}
where we wrote $\veps_2\veps_3\sgn(z_{23})=\sgn(z_{12}z_{31}z_{23})$. This correlator is obtained from a dimension six operator proportional to $F^3$.

Finally, for the last correlator we will use the projector $\partial_{\zb}X^A_1,\partial_{\zb}X^B_2,\partial_{z}X^C_3$ to obtain the correlator between fields of helicity $-1,-1,+1$. The correlator is of the form
\begin{align}
  \label{eq:finalspin111}
\langle \mathcal{P}^{(\veps_1,\veps_2,\veps_3)}[\phi^{(\Delta_1,1)}_{\zb}\phi^{(\Delta_2,1)}_{\zb}\phi^{(\Delta_3,1)}_{z}]\rangle&=\frac{-2\veps_2\veps_3(2\pi)^2C_{3}z^2_{12}}{z_{23}|z_{12}||z_{31}|}
            \left(\frac{\epsilon_3z_{23}}{\epsilon_1z_{12}}\right)^{1-\Delta_3}\left(\frac{\epsilon_2z_{23}}{\epsilon_1z_{31}}\right)^{1-\Delta_2} \nonumber
  \\\delta(\zb_{13})\delta(\zb_{23}) &\Theta\left(\frac{\epsilon_2z_{23}}{\epsilon_1z_{31}}\right) \Theta\left(\frac{\epsilon_3z_{23}}{\epsilon_1z_{12}}\right)\,
   \frac{\Gamma(\Delta-3)}{\left(\frac{\veps_1}{z_{23}}(u_1z_{23}+u_2z_{31}+u_3z_{12})\right)^{\Delta-3}}\,,
\end{align}
In the limit $\Delta_i\to 1$ we reproduce the well-known three-point MHV amplitude for color-ordered gluons written in an asymptotic momentum basis; cf. equation \eqref{eq:43}
\begin{align}
  \label{eq:gluonMHV}
\langle \mathcal{P}^{(\veps_1,\veps_2,\veps_3)}[\phib^{(1)}_{\zb}\phib^{(1)}_{\zb}\phib^{(1)}_z]\rangle&=\sgn(z_{12}z_{23}z_{31})\frac{-2(2\pi)^2C_{3}|z_{12}|}{|z_{23}||z_{31}|}
  \delta(\zb_{13})\delta(\zb_{23})\nonumber
  \\&\Theta\left(\frac{\epsilon_2z_{23}}{\epsilon_1z_{31}}\right) \Theta\left(\frac{\epsilon_3z_{23}}{\epsilon_1z_{12}}\right)\,
   \lim_{\Delta\to 3}\frac{\Gamma(\Delta-3)}{\left(\frac{\veps_1}{z_{23}}(u_1z_{23}+u_2z_{31}+u_3z_{12})\right)^{\Delta-3}}\,.
\end{align}

\section{Discussion and outlook}
\label{sec:discussion}
In this work we developed an embedding space formalism for carrollian CFTs and employed it to determine two- and three-point functions of spinning fields. Alternatively, one can view the results presented herein as computing two- and three-point S-matrix elements in a basis of asymptotic position states. From this perspective, the presented approach is clearly less economical than using the little group method in momentum space to determine three point functions.  Nevertheless, this approach potentially opens the door to a deeper analysis of conformal carrollian theories along the following lines.

\paragraph{Generalizations to higher spin fields.} In the present work, we restricted ourselves to fields of spin zero and spin one. However, it is straightforward to apply the present approach also to fields of spin two and higher. The main difficulty is to determine the tensors structures $W_{A_1\ldots A_{s}B_1\ldots B_s}(X_1,X_2)$ and $W_{A_1\ldots A_{s_1}B_1\ldots B_{s_2}C_1\ldots C_{s_3}}(X_1,X_2,X_3)$ appearing in two- and three-point functions, respectively, subject to the conditions \eqref{eq:transverscon} and \eqref{eq:defdedon}. An efficient approach to compute these tensors for conventional CFTs was developed in the work \cite{Costa:2011mg}. In this work, fields $\Phi_{A_1\ldots A_s}$ are encoded in polynomials $\Phi_{A_1\ldots A_s} Z^{A_1}\ldots Z^{A_s}$ which allows to write a basis of building blocks for spinning correlators as
\begin{align}
  \label{eq:Qbuild}
  V_{i,jk}&=\frac{(Z_i\cdot X_j)(X_i\cdot X_k)-(Z_i\cdot X_k)(X_i\cdot X_j)}{X_{jk}}\\
  H_{ij}&=-2[(Z_i\cdot Z_j)(X_i\cdot X_j)-(Z_i\cdot X_j)(Z_j\cdot X_i)]\,.
\end{align}
The index structure can be recovered by acting $(s_1+s_2+s_3)$-times with the operator
\begin{equation}
  \label{eq:spinrec}
  D_A=\left (1+Z \cdot \frac{\partial}{\partial Z}\right) \frac{\partial}{\partial Z^A}
    -\frac{1}{2}Z_A\frac{\partial^2}{\partial Z\cdot \partial Z}.
  \end{equation}
  Acting with these operator produces a correlator that is transverse with respect to $X^A$ and traceless by construction. In order to obtain a correlator for fields obeying $I^A$, it is enough to act with the projector \eqref{eq:deftransproj} on each index.

  The number of independent tensors for CFT n-point functions was determined in \cite{Costa:2011mg}. We can adopt this for our approach keeping in mind that the preferred point $I^A$ is also available for constructing the tensors. One therefore finds that the number of independent tensors for a two-point (three-point) function for a carrollian CFT in the embedding space is given by the number of independent tensors for a \emph{three-point (four-point)} function in a conventional CFT. Thus, one finds for the two-point function of a carrollian CFT that the tensor is given by
  \begin{equation}
  \label{eq:Qlincomb}
  \prod_i V^{m_i}_{i,kl}\prod_{i<j} H^{n_{ij}}_{ij}\qquad m_i+\sum_{j\neq i}n_{ij}=l_i
\end{equation}
where $X^A_i=X^A_1,X^A_2,I^{A}$. Similarly, the tensor structure for a carrollian three-point function can be obtained as a linear combination of
\begin{equation}
  \label{eq:three-point-tensorstr}
\prod_iV^{m_i}_{i;2I}\prod_iV^{\bar{m}_i}_{i;3I}\prod_{i<j}H^{n_{ij}}_{ij}\qquad m_i+\bar{m}_i+\sum_{j\neq i}n_{ij}=s_i\,
\end{equation}
with the coefficients being arbitrary functions of $X_1,X_2,X_3,I$ that obey the scaling. The number of independent structures is given by the set of all positive integers such that $n_{12}+n_{13}\le s_1,n_{12}+n_{23}\le s_2,n_{13}+n_{23}\le s_3$.

The condition \eqref{eq:defdedon} can also be imposed efficiently in the indexless framework and the adaptation to the present context should be straightforward. A generalization of our approach to include fields of mixed symmetry or half-integer spin should follow along the lines of \cite{Elkhidir:2014woa,Costa:2014rya}. It would be interesting to work out the details of these generalizations.

\paragraph{Techniques from conventional CFTs.}
One of the aims of using the embedding space approach to compute conformal carrollian correlators was to phrase carrollian CFTs in a way that might allow the adaptation of techniques from conventional CFTs. For instance, we saw above that the ansatz for two- and three-point functions of carrollian CFTs take a form of a sum over three- and four-point functions in conventional CFTs.\footnote{Note, however, that this depends on the assumption that the corresponding functions in \eqref{eq:caransatz} and the three-point ansatz are analytic. It is not expected that this holds beyond tree level.} It would be interesting to see whether this similarity can be pushed further. Moreover, the embedding space approach lends itself to a description of conformal blocks or the OPE, e.g., \cite{Costa:2011dw,Fortin:2016lmf}. At the moment it is not clear if and how these crucial properties of usual CFTs are realized in carrollian CFTs. However, it is likely that the present embedding space approach will simplify the analysis of these matters.

\paragraph{Another point of view: working directly on $\scri$.}
The prescription \eqref{eq:centralchain} that we use to obtain the conformal carrollian boundary field refers explicitly to a four-dimensional field $\phi_{\mu_1\ldots \mu_s}(x^\mu)$ in Minkowski space. Correspondingly, the limit $\sigma \to 0$ that we take to obtain the conformal carrollian correlators from the embedding space approach can be viewed as pushing a generic Poincaré-invariant correlator to the boundary. We note that there is an equivalent approach to extract the carrollian correlators from the embedding space correlator that is more intrinsic to $\scri$, which we want to sketch briefly.

Consider the embedding space $\mathbb{R}^{1,5}$ for a Euclidean CFT and choose the parametrization
\begin{equation}
  \label{eq:13}
  X_E=\sinh \tau\left(\sinh \tau, \sin \psi n^i,\frac{1}{\sqrt{2}}(\cosh \tau+\cos \psi),\frac{1}{\sqrt{2}}(-\cosh \tau+\cos \psi)\right)
\end{equation}
for the Poincaré section of the light-cone $X^2_E=0$ with $n^in^j\delta_{ij}=1$.
Here $\tau$ denote Euclidean time. Continuing this now to Lorentzian
signature by setting $\tau=it+\epsilon$ and linearizing in
$\epsilon$ (and using $X_{E}^{0}=iX^{0}$ where the other ones stay
inert) leads us to
\begin{align}
  X&=\left(
         \sin t ,
         \sin \psi n^i, \frac{1}{\sqrt{2}}(-\cos t+\cos \psi),
         \frac{1}{\sqrt{2}}(\cos t+\cos \psi)
         \right)  \nonumber \\
       &\quad
        +i \epsilon \left(
         -  \cos t, 0,0,0,
                  -\sin t,
          \sin t,
         \right)\,.
         \end{align}
This is a parametrization for the Lorentzian cylinder, the conformal compactification of Minkowski space. The surface $\scri^+$ in these coordinates is reached by setting $t=\pi-\psi$ so that we find the parametrization
\begin{equation}
  \label{eq:alternativeX}
  X= \left(\frac{q^{\mu}}{1+z\zb}, u,0\right) + i \epsilon \left(u,0,0,0,-1,1\right)\,,
\end{equation}
where we set $\psi=\cot^{-1}(u/\sqrt{2})$, rescaled by $\sin \psi$ and used stereographic coordinates on the sphere. An equivalent prescription leads to a parametrization of $\scri^-$.
If we use \eqref{eq:alternativeX}, and its equivalent on $\scri^-$, instead of \eqref{eq:82} in the ansatz \eqref{eq:caransatz} for conformal carrollian fields, the correlator \eqref{eq:finalBagchi} arises in the limit $\epsilon_1,\epsilon_2\to 0$.\footnote{In order to regulate the expressions and ensure the convergence of the Schwinger parametrization, one could set  $t=\pi-\psi+\delta$ with $\delta\to 0$.}
This approach should be equivalent to the one adopted in the present work, but appears to be closer to the works \cite{Herfray:2020rvq,Herfray:2021xyp,Herfray:2021qmp,Bekaert:2022oeh}. The present approach was chosen in order to make the link to the analysis of \cite{Donnay:2022aba,Donnay:2022wvx} explicit.

\paragraph{Generalization to non-trivial backgrounds.}
CFTs are usually considered on conformally flat backgrounds. However, it is clearly possible to consider CFTs on non-trivial backgrounds, e.g., in the context of the AdS/CFT correspondence by considering a boundary metric that is not conformally flat or in the closely related ambient space approach; cf. \cite{Parisini:2022wkb}.

The works \cite{Herfray:2020rvq,Herfray:2021xyp,Herfray:2021qmp} developed a framework to describe null infinity of a general asymptotically flat spacetime in terms of tractor geometry. In particular, the presence of gravitational radiation is encoded in the (non-unique) choice of connection compatible with the conformal carrollian structure; see also \cite{Ashtekar:1981hw} for an earlier account of this. It would be very interesting to see whether the present approach can be generalized along the above lines to allow the computation of conformal carrollian correlators on non-trivial backgrounds. In this way, one would obtain carrollian CFTs defined on the null infinity of asymptotically flat spacetimes with outgoing gravitational radiation.

\paragraph{Interpretation of the two branches.}
As was observed before, conformal carrollian correlators exhibit two branches: a $u$-independent branch belonging to the magnetic carrollian sector and a $u$-dependent branch corresponding to the electric theory. We found that low-point correlators in the latter are distributionally valued. According to the proposal of \cite{Donnay:2022aba,Donnay:2022wvx,Bagchi:2022emh} it is the electric sector that is related to scattering in asymptotically flat spacetimes via the maps presented in Section \ref{sec:Smatrix}. The interpretation of the magnetic branch is less clear. One possibility is that these modes correspond to the Goldstone modes associated with the infinite-dimensional symmetries of massless gauge theories in flat spacetimes. The fact that effective actions for these Goldstone modes that have been constructed in the literature are of the magnetic type \cite{Barnich:2017jgw,Barnich:2012rz,Barnich:2022bni,Merbis:2019wgk} or essentially two-dimensional \cite{Nguyen:2020hot,Nguyen:2021ydb,Kalyanapuram:2021bvf,Kapec:2021eug} would suggest this interpretation.

\paragraph{Flat space holography and AdS/CFT.}
The relation between conformal carrollian field theories as a holographic dual to scattering in asymptotically flat spacetimes  is intriguing but still in its infancy. In this work we took a step towards establishing such a correspondence by showing that carrollian three-point functions (of low spin) coincide with three-point amplitudes in Minkowski space. However, it can still be argued that both sides of this correspondence are fixed by Poincaré symmetry. In order to see non-trivial evidence for such a correspondence one would need a simplified model in which one is able to independently define both the S-matrix of a bulk theory in asymptotically flat spacetime and a carrollian CFT on the boundary. A possible check would then consist in computing four-point functions in both theories.

The clearest way to such a set-up appears to be a suitable limit on both sides of an established duality in the framework of the AdS/CFT correspondence. The idea of obtaining insight about asymptotically flat spacetimes from such a limit goes back to \cite{Polchinski:1999ry,Susskind:1998vk,Giddings:1999jq}. The link between S-matrix amplitudes and the large radius limit of AdS were discussed in \cite{Penedones:2010ue,Fitzpatrick:2011hu,Raju:2012zr,Paulos:2016fap,Li:2021snj,Hijano:2020szl}. More recently, the flat space limit of $\textrm{AdS}_4/\textrm{CFT}_3$ was related to celestial holography \cite{deGioia:2022fcn,deGioia:2023cbd} and carrollian CFTs \cite{Bagchi:2023fbj} (see also \cite{Hijano:2019qmi} for related work in three dimensions).
We want to conclude by briefly commenting on the way the present approach fits into this discussion. The embedding space $\mathbb{R}^{4,2}$ not only allows to embed the conformal light-cone $X^2=0, X\sim \lambda X$ but also $\textrm{AdS}_5$ as $X^2=-\frac{1}{\ell^2}$. In the context of the AdS/CFT correspondence it is clear that one thinks of the former as the boundary of the latter. In our choice of coordinates, the surface $I\cdot X=\sigma=0$ singles out a null-surface on the boundary extending into the bulk. On the other hand, for fixed $\sigma\neq 0$ the surface $I\cdot X=\sigma$ is time-like in the bulk with the induced metric being Minkowski space. Our prescription \eqref{eq:centralchain} then corresponds to defining the correlator on a family of such surfaces and subsequently pushing the surface towards the null surface $\sigma=0$ while simultaneously going out towards the boundary of $\textrm{AdS}_5$. It would be very interesting to see if anything can be learned about three-dimensional carrollian CFTs, and the presumed duality to scattering in asymptotically flat spacetimes, by considering such a set-up in the context of AdS/CFT.

\begin{acknowledgments}
 The author thanks Arjun Bagchi, Adrien Fiorucci, Oscar Fuentealba, Hernán González, Enrico Parisini, Alfredo Pérez, and Romain Ruzziconi for discussions and an anonymous referee for comments that led to improvements of the manuscript.
  He is also grateful to José Figueroa-O'Farrill, Emil
  Have, and Stefan Prohazka for collaboration on celestial and carrollian matters.
  The author thanks Stefan Prohazka in particular for collaboration during intermediate stages of this project.

  This work was supported by a Marina Solvay-fellowship and
  the F.R.S.-FNRS Belgium through the convention IISN 4.4503. The author thanks the organizers of the ``Quantum Gravity around the Corner'' workshop at Perimeter Institute, Luca Ciambelli and Céline Zwikel, where preliminary results of this project were presented. He also thanks CECS, Valdivia for hospitality during the final stages of this work.
\end{acknowledgments}

\appendix

\newpage
\section{Conventions and useful equations}
\label{sec:embcoord}
Minkowski space in flat null coordinates takes the form
\begin{equation}
  \label{eq:flatbondi}
  \dd s^2=-2\dd u \dd r +2 r^2 \dd z \dd \zb\,.
\end{equation}
In terms of Cartesian coordinates these coordinates can be written as
\begin{equation}
  \label{eq:8}
  x^\mu=u n^\mu+r q^\mu\,,
\end{equation}
where
\begin{equation}
  \label{eq:38}
  n^\mu=\frac{1}{\sqrt{2}}\begin{pmatrix}1\\0\\0\\-1\end{pmatrix}\qquad q^\mu=\frac{1}{\sqrt{2}}\begin{pmatrix}1+z\zb\\z+\zb\\-i(z-\zb)\\1-z\zb\end{pmatrix}\,.
\end{equation}
We have the following relations
\begin{equation}
  \label{eq:56}
  n^\mu q_\mu=-1\qquad q^\mu(z_1,\zb_1)q_\mu(z_2,\zb_2)=-|z_{12}|^2\qquad x^\mu x_\mu=-2ur
\end{equation}
The inner product of two points on the light cone $X^{2}=0$ in the parametrization \eqref{eq:82} is given
by
\begin{align}
  \label{eq:XidotXj}
  X_{ij} = X_{i} \cdot X_{j}
  = \sigma_{i}\sigma_{j}
  \left(r_{ij} u_{ij} - r_{i}r_{j} |z_{ij}|^{2} \right)
\end{align}
where $r_{ij}= r_{i}-r_{j}$, $u_{ij}=u_{i}-u_{j}$ and
$|z_{ij}|^2 = (z_{i}-z_{j})(\zb_{i}-\zb_{j})$.
Upon using our prescription of Section~\ref{eq:correlators} we set
 $r_{i}= \veps_{i} \sigma^{-1}$ which leads
to ($\veps_{ij} = \veps_{i} - \veps_{j}$)
\begin{align}
  X_{ij} = (\veps_i\sigma_j-\veps_j\sigma_i) u_{ij} - \veps_{i}\veps_{j} |z_{ij}|^{2} \, .
\end{align}
We also note the relations
\begin{equation}
  \label{eq:64}
  \partial_{\zb_i}X_i\cdot X_j=-\veps_i\veps_jz_{ij}\qquad \partial_{z_i}X_i\cdot X_j=-\veps_i\veps_j\zb_{ij}\qquad \partial_{\zb_i}X_i\cdot \partial_{z_j}X_j=\veps_i\veps_j\,.
\end{equation}
\section{Symmetry transformations in embedding space}
\label{sec:syminembed}
Consider the following parametrization for points $X^A$ in the embedding space
\begin{equation}
  \label{eq:embedcoord}
X^A=\begin{pmatrix}X^\mu\\X^+\\X^-\end{pmatrix}=\sigma\begin{pmatrix}x^\mu\\-\frac{x^2}{2}+\frac{\Lambda}{2\sigma^2}\\1\end{pmatrix} \qquad x^\mu=u n^\mu+r q^\mu(z,\zb)\,.
\end{equation}
We note the relation $X^Ag_{AB}X^B=\Lambda$ so that constant $\Lambda$ denotes a particular quadric in the embedding space. For $\Lambda=0$ we reproduce the parametrization \eqref{eq:82} in the main text.
The Killing vector fields leaving the metric \eqref{eq:31} invariant are given by
\begin{equation}
  \label{eq:71}
  M_{AB}=X_A\partial_B-X_B\partial_A\,.
\end{equation}
Using
\begin{subequations}
 \begin{align}
   \frac{\partial}{\partial X^\mu}&=\frac{1}{\sigma}\partial_\mu+2\sigma x_\mu\partial_\Lambda\\
   \frac{\partial}{\partial X^+}&=2\sigma \partial_\Lambda\\
   \frac{\partial}{\partial X^-}&=-\frac{x^\mu}{\sigma}\partial_\mu+\partial_\sigma+2\sigma(-\frac{x^2}{2}+\frac{\Lambda}{2\sigma^2})\partial_\Lambda
 \end{align}
 \end{subequations}
 we can write these as
 \begin{subequations}
    \begin{align}
   M_{+\mu}&=\partial_\mu\\
      M_{\mu\nu}&=x_\mu\partial_\nu-x_{\nu}\partial_\mu\\
       M_{+-}&=\sigma\partial_\sigma+x^\mu\partial_\mu\\
      M_{-\mu}&=x_\mu x^\lambda\partial_\lambda+(-\frac{x^2}{2}+\frac{\Lambda}{2\sigma^2})\partial_\mu-\sigma x_\mu\partial_\sigma\,.
 \end{align}
\end{subequations}
In terms of the four-dimensional conformal algebra we can recognize these transformations as translations, Lorentz transformations, dilations, and special conformal transformations, respectively. Note that all transformations leave surfaces of constant $\Lambda$ invariant and the first two leave surfaces of constant $\sigma$ invariant. We also note that the Euler vector field
\begin{equation}
  \label{eq:75}
  X^A\partial_A=2\Lambda\partial_\Lambda+\sigma\partial_\sigma\,,
\end{equation}
becomes $X^A\partial_A=\sigma \partial_\sigma$ on the lightcone $\Lambda=0$.

With our choice of parametrization $x^\mu=u n^\mu+r q^\mu$ we can further write for translations and Lorentz transformations
\begin{align}
  \label{eq:74}
  M_{+\mu}&=\partial_\mu=-q_\mu\partial_u-n_\mu\partial_r+r^{-1}\partial q_\mu \bar{\partial}+r^{-1}\bar{\partial} q_\mu \partial\\
M_{\mu\nu}&=x_\mu\partial_\nu-x_\nu\partial_\mu=2(-n_{[\mu}q_{\nu]}(u\partial_u-r\partial_r)+ur^{-1}n_{[\mu}\bar{\partial}q_{\nu]}\partial+q_{[\mu}\bar{\partial}q_{\nu]}\partial+c.c.)\,.
\end{align}

We can now use these expressions to show that a field $\Phi_{A_1\ldots A_s}(X)$  of scaling dimension $\Delta$ in the embedding space transforms indeed like a conformal carrollian quasi-primary field when evaluated on $r\sigma=1$ in the limit $\sigma\to 0$ and projected using $\partial_z X^{A_1}$. To avoid a proliferation of indices we will restrict ourselves to the case of a spin one field. Starting with translations we have the Killing vector $\xi=T^\mu M_{+\mu}$ and therefore
\begin{equation}
  \label{eq:76}
  \mathcal{L}_\xi\Phi_{A_1}=T^{\mu}M_{+\mu}\Phi_{A_1}+\Phi_{\mu}\partial_{A_1}T^\mu=T^{\mu}M_{+\mu}\Phi_{A_1}\,,
\end{equation}
since $T^\mu$ is constant. We have then
\begin{align}
  \label{eq:77}
  \delta_T\phi^{(\Delta,1)}_z&=\lim_{\sigma\to 0}\partial_zX^{A_1}T^{\mu}M_{+\mu}\Phi_{A_1}|_{r\sigma=1}\\
                                             &=\lim_{\sigma\to 0}\partial_zX^{A_1}T^\mu(-q_\mu\partial_u-\Delta\sigma  q_\mu+\sigma \partial q_\mu \bar{\partial}+\sigma \bar{\partial} q_\mu\partial )\Phi(X)_{A_1}|_{r\sigma=1}\\
  &=(-T^\mu q_\mu\partial_u)\phi^{(\Delta,1)}_z\,,
\end{align}
since all other terms vanish in the limit. Setting $-T^\mu q_\mu=\mathcal{T}(z,\zb)$, a linear combination of $1,z,\zb,z\zb$, this is indeed the transformation law of a conformal carrollian quasi-primary under translations.

For Lorentz transformations we have the Killing vector $\xi=a^{\mu\nu}M_{\mu\nu}$ with $a^{\mu\nu}$ antisymmetric for which we find
\begin{align}
  \label{eq:78}
  \partial_zX^{A_1}&\mathcal{L}_\xi \Phi_{A_1}=\partial_zX^{A_1}a^{\mu\nu}X_{\mu}\frac{\partial}{\partial X^\nu}\Phi_{A_1}+\Phi_{\nu}a\indices{^\nu_{\mu}}\delta^{\mu}_{A_1}\partial_zX^{A_1}\, \nonumber \\
                                             &=\partial q^\lambda a^{\mu\nu}(-n_{\mu}q_{\nu}(u\partial_u-r\partial_r)+ur^{-1}n_{\mu}\bar{\partial}q_{\nu}\partial+q_{\mu}\bar{\partial}q_{\nu}\partial+c.c.)\Phi_{\lambda}-\Phi_{\nu}a\indices{^\nu_{\lambda}}\partial q^\lambda \nonumber \\
  &\overset{r\sigma=1}{=}\partial q^\lambda a^{\mu\nu}(-n_{\mu}q_{\nu}(u\partial_u+\sigma \partial_\sigma)+u \sigma n_{\mu}\bar{\partial}q_{\nu}\partial+q_{\mu}\bar{\partial}q_{\nu}\partial+c.c.)\Phi_{\lambda}- \Phi_{\nu}a\indices{^\nu_{\lambda}}\partial q^\lambda \nonumber \\
                   &=a^{\mu\nu}(-n_{\mu}q_{\nu}(u\partial_u-\Delta)+u \sigma n_{\mu}\bar{\partial}q_{\nu}\partial+q_{\mu}\bar{\partial}q_{\nu}\partial+c.c.)(\partial q^\lambda\Phi_{\lambda}) \\
  &\qquad \qquad -a^{\mu\nu}q_{\mu}\partial q_{\nu}n^{\lambda}\Phi_{\lambda}- \Phi_{\nu}a\indices{^\nu_{\lambda}}\partial q^\lambda\nonumber
\end{align}
Using completeness of the basis spanned by $n^\mu,q^\mu,\partial q^\mu,\bar{\partial}q^\mu$ we can write
\begin{equation}
  \label{eq:79}
\Phi_\nu a\indices{^\nu_\lambda}\partial q^\lambda=\Phi_\nu(-n^\nu q_\mu-q^\nu n_\mu+\partial q^\nu\bar{\partial}_\mu q)a\indices{^\mu_\lambda}\partial q^\lambda\,,
\end{equation}
where we used antisymmetry of $a^{\mu\nu}$. Taking into account the gauge-fixing conditions
\begin{equation}
  \label{eq:83}
  I^A\Phi_A=0\Leftrightarrow \Phi_+=0\qquad X^A\Phi_A=(\sigma u n^\mu+\sigma r q^\mu)\Phi_\mu-\frac{\sigma x^2}{2}\Phi_++\sigma \Phi_-=0\rightarrow q^\mu\Phi_\mu=0\,,
\end{equation}
the limit $\sigma \to 0$ in \eqref{eq:78} yields
\begin{align}
  \label{eq:81}
  \lim_{\sigma \to 0}\partial_zX^{A_1}\mathcal{L}_\xi \Phi_{A_1}|_{r\sigma=1}=a^{\mu\nu}(-n_\mu q_\nu u\partial_u+\Delta q_\mu n_\nu+q_\mu \bar{\partial} q_\nu \partial+\partial q_\mu \bar{\partial}q_\nu+q_\mu \partial q_\nu \bar{\partial})\phi^{(\Delta,1)}_z\,.
\end{align}
Setting $a^{\mu\nu}q_{\mu}\bar{\partial}q_\nu=Y(z)$ and $a^{\mu\nu}q_{\mu}\partial q_\nu=\bar{Y}(\zb)$, which are both polynomials of order two in $z$ and $\zb$, respectively, we can write this as
\begin{align}
  \label{eq:84}
  \lim_{\sigma \to 0}\partial_zX^{A_1}\mathcal{L}_\xi \Phi_{A_1}|_{r\sigma=1}&=\left(\frac{1}{2}(\partial Y+\bar{\partial}\bar{Y})(u\partial_u+\frac{\Delta}{2})+\frac{1}{2}(\partial Y-\bar{\partial}\bar{Y})+Y\partial+\bar{Y}\bar{\partial}\right)\phi^{(\Delta,1)}_z\\
&=\left(\frac{1}{2}(\partial Y+\bar{\partial}\bar{Y})u\partial_u+Y\partial+\bar{Y}\bar{\partial}+\frac{\Delta+1}{2}\partial Y+\frac{\Delta-1}{2}\bar{\partial} \bar{Y}\right)\phi^{(\Delta,1)}_z\,.
\end{align}
This is indeed the transformation law of a conformal carrollian quasi-primary field.

\section{Scattering amplitudes in position space}
\label{sec:scatt-ampl-moment}

In order to compare with the expressions derived in the main text, we
collect in this appendix some well-known expressions for low-point
scattering amplitudes in momentum space converted to position space amplitudes using the integral transform discussed in Section \ref{sec:Smatrix} the derivation of which we briefly review in the case of a scalar field; for the spinning case see \cite{Donnay:2022wvx}.

Let $\phi(x)$ be a scalar field in Minkowski space with coordinates \eqref{eq:flatbondi}. In term of usual plane-wave modes the scalar field can be expanded as
\begin{equation}
  \label{eq:4}
  \phi(x)=\int\frac{\dd^3k}{(2\pi)^32k^0}(a_ke^{ik\cdot x}+a^\dagger_ke^{-ik\cdot x})\,.
\end{equation}
Evaluating this expression with the usual parametrization $k=\omega q^\mu(w,\bar{w})$ in the limit $r\to \infty$ (assuming $\omega>0$) one obtains
\begin{equation}
  \label{eq:7}
  \phi(x)=\frac{-i}{8\pi^2}r^{-1}\int^\infty_0\dd \omega(a_{\omega q} e^{-i\omega u}-a^\dagger_{\omega q}
  e^{i\omega u})\,,
\end{equation}
from which one defines the carrollian boundary field
\begin{equation}
  \label{eq:15}
  \phib(u,z,\zb)\equiv \lim_{r\to \infty}r \phi(x)=\frac{-i}{8\pi^2}\int^\infty_0\dd \omega(a_{\omega q} e^{-i\omega u}-a^\dagger_{\omega q}  e^{i\omega u})\,.
\end{equation}
The above relation between creation and annihilation operators $a,a^\dagger$ on the one hand and $\phib(u,z,\zb)$ on the other hand leads to the definition
\begin{equation}
  \label{eq:20}
\ket{u,z,\zb}=\frac{1}{2\pi}\int^\infty_0\dd \omega e^{i\omega u}\ket{\omega,z,\zb}\,,\qquad  \bra{u,z,\zb}=\frac{1}{2\pi}\int^\infty_0\dd \omega e^{-i\omega u}\bra{\omega,z,\zb}\,,
\end{equation}
of asymptotic position states $\ket{u,z,\zb}$ in terms of usual momentum states $\ket{\omega,z,\zb}$ created by $a^\dagger_k$. Applying the transformation to all asymptotic states leads to the integral transform \eqref{eq:65}.

The transformation \eqref{eq:7} is invertible under the assumption $\omega>0$
\begin{equation}
  \label{eq:21}
  a_{\omega q}=4\pi i\int^\infty_{-\infty}\dd u e^{i\omega u}  \phib(u,z,\zb),\qquad a^\dagger_{\omega q}=-4\pi i\int^\infty_{-\infty}\dd u e^{-i\omega u}  \phib(u,z,\zb).
\end{equation}
Plugging this form of the operators back into \eqref{eq:4} and freely changing the order of integration yields
\begin{equation}
  \label{eq:29}
  \phi(x)=\frac{1}{4\pi^2}\int \dd^2w\dd u \left[ i\int^\infty_0\dd \omega \omega e^{i \omega(q\cdot x+u)}\phib(u,z,\zb)-i\int^\infty_0\dd \omega \omega e^{-i \omega(q\cdot x+u)}\phib(u,z,\zb)\right]\,.
\end{equation}
This can be interpreted as an expansion of the scalar field in terms of modes
\begin{equation}
  \label{eq:30}
  \psi(u,z,\zb;x)=\lim_{\epsilon\to 0}i\int^\infty_0\dd \omega\, \omega\, e^{i \omega(q\cdot x+u+i\epsilon)}=\lim_{\epsilon\to 0}\frac{-i}{(u+q\cdot x+i\epsilon)^2}\,,
\end{equation}
which solve the free wave equation.

Combining the two terms in  \eqref{eq:29} and integrating by parts one finds
\begin{align}
  \label{eq:32}
  \phi(x)&=\frac{1}{2\pi}\int \dd^2w\dd u\partial_u\delta(u+q\cdot x)\phib(u,z,\zb)\nonumber\\
  &=-\frac{1}{2\pi}\int \dd^2w\,\partial_u\phib(u=-q\cdot x,z,\zb)\,,
\end{align}
the Kirchhoff-d'Adhémar formula \cite{Penrose:1985bww}. We see therefore that the $u$-derivative of the carrollian field $\partial_u\phib(u,z,\zb)$ is in fact enough to recover the whole field in the bulk.\footnote{The author thanks Romain Ruzziconi for conversations on this issue.} This comes at the expense of loosing track of the zero modes of $\phib(u,z,\zb)$. However, due to the assumption $\omega>0$ in the beginning, these modes need a separate treatment in any case. In the gravitational case, the field $\phib$ corresponds to the shear $C_{zz}$ while $\partial_u\phib$ corresponds to the news $N_{zz}$.

Let us now apply this framework to two- and three-point amplitudes.

The 2-point scattering amplitude is equal to the inner product of
one-particle states on an equal time-slice. For a massless particle of
momentum $p^\mu=\omega q^\mu$ with $q^\mu$ as in \eqref{eq:82} and helicity $h$ this is given by
\begin{equation}
  \label{eq:2pointmom}
\braket{ p_2,h_{2} | p_1,h_1}=(2\pi)^3\frac{\delta(\omega_1-\omega_2)}{\omega_1}\delta(z_{12})\delta(\zb_{12}) \delta_{h_1,-h_2}\,.
\end{equation}
We will take this to be the two-particle amplitude $A_2(\omega_2,z_2,\zb_2,\veps_1=+1,\omega_1,z_1,\zb_1,\veps=-1)$, where $\veps=+1(-1)$ denotes an outgoing (incoming) particle. Note that we label all helicities as if they belong to outgoing particles.

To switch to the asymptotic position space basis, we apply the integral transformation
\begin{equation}
  \label{eq:68}
  \mathcal{C}_2(u_2,z_2,\zb_2;u_1,z_1,\zb_1)=\frac{1}{(2\pi)^2}\int^\infty_0\dd \omega_1\dd \omega_2 e^{-i\veps_1\omega_1 u_1}e^{-i\veps_2\omega_2 u_2}\mathcal{A}_2\,.
\end{equation}
In the present case this leads to
\begin{equation}
  \label{eq:72}
 \mathcal{C}_2(u_2,z_2,\zb_2;u_1,z_1,\zb_1)=(2\pi)\int^\infty_0\frac{\dd \omega}{\omega}e^{-i\omega(u_2-u_1)}\delta(z_{12})\delta(\zb_{12}) \delta_{h_1,-h_2}\equiv I(1)\,.
\end{equation}
This integral does not converge. However, defining $I(1)=\lim_{\epsilon\to 0,\Delta \to 1}I_\epsilon(\Delta)$ with
\begin{equation}
  \label{eq:73}
I_\epsilon(\Delta)=\int^\infty_0\frac{\dd \omega}{\omega}\omega^{\Delta-1} e^{-i\omega(u_2-u_1)-\omega \epsilon}=\frac{i^{\Delta-1}\Gamma(\Delta-1)}{(u_{21}-i\epsilon)^{\Delta-1}}
\end{equation}
we find
\begin{equation}
  \label{eq:85}
   \mathcal{C}_2(u_2,z_2,\zb_2;u_1,z_1,\zb_1)=(2\pi)\delta(z_{12})\delta(\zb_{12}) \delta_{h_1,-h_2}\lim_{\Delta\to 1,\epsilon\to 0}\frac{i^{\Delta-1}\Gamma(\Delta-1)}{(u_{21}-i\epsilon)^{\Delta-1}}\,.
 \end{equation}
 This is indeed of the form of the two-point correlator presented in Section \ref{eq:carcor2pt}.

 In the limit $\Delta\to 1$ this correlator is divergent and of the form
 \begin{equation}
   \label{eq:2} \mathcal{C}_2=\frac{C'}{\Delta-2}\delta(z_{12})\delta(\zb_{12})+C'(\gamma_E+\ln|\veps_{12}u_{12}|-i\pi\Theta(-\veps_{12}u_{12}))\delta(z_{12})\delta(\zb_{12})\,.
 \end{equation}
 However, as seen from equation \eqref{eq:32}, it is the $u$-derivative of the carrollian field that encodes the necessary data to reconstruct the bulk field. The full carrollian field $\phib$ contains data about the zero-modes that need a separate treatment. Consistent with this, the divergent piece in \eqref{eq:2} is $u$-independent. The correlator of two ``news fields'' $\partial_u\phi$ is finite and given by
 \begin{equation}
   \label{eq:33}
   \langle \partial_{u_2}\phi(u_2,z_2,\zb_2)\partial_{u_1}\phi(u_1,z_1,\zb_1) \rangle=C'(u^{-2}_{21}-i\pi \delta^{(1)}(u_{21}))\,.
 \end{equation}
 It is these finite correlators that are physically relevant. We will nevertheless consider correlators of the carrollian fields $\phib$ in the following since the physically relevant correlators can be reconstructed from the former.\footnote{The divergence in the correlator of the fields $\phib$ can be traced back to the fact that the modes \eqref{eq:30} are not delta function-normalizable.}

 Let us turn to three-point amplitudes.
Due to 3-particle special kinematics, on-shell amplitudes for massless fields vanish in Lorentz signature. In order to find non-vanishing amplitudes, we will switch to $(2,2)$-signature Minkowski space or, more generally, complex momenta. Lorentz invariance then fixes all color-stripped 3-particle amplitudes (up to a constant factor), cf., e.g., \cite{Benincasa:2007xk,Arkani-Hamed:2017jhn}. Labelling all helicities as outgoing, the resulting amplitude for fields with helicities $h_1,h_2,h_3$ is most conveniently displayed in the spinor-helicity formalism where one finds
\begin{align}
  \label{eq:spinhel3pt}
  A_3(1^{h_1}2^{h_2}3^{h_3})=\!\!\begin{cases}c \braket{12}^{h_3-h_1-h_2}\braket{31}^{h_2-h_1-h_3}\braket{23}^{h_1-h_2-h_3}\delta^{(4)}(\sum_ip_i),\!\!\! &h_1+h_2+h_3<0\\
    \tilde{c}\, [12]^{-h_3+h_1+h_2}[31]^{-h_2+h_1+h_3}[23]^{-h_1+h_2+h_3}\delta^{(4)}(\sum_ip_i),\!\!\! &h_1+h_2+h_3>0.\end{cases}
\end{align}
For momenta of the form $p^\mu=\veps \omega q^\mu$ with $\veps=\pm 1$ for outgoing/ingoing particles, we choose spinor helicity variables such that $\langle ij\rangle=\sqrt{\omega_i\omega_j}z_{ij}$ and $[ij]=-\veps_i\veps_j\sqrt{\omega_i\omega_j}\zb_{ij}$. The above then translates to
\begin{align}
  \label{eq:41}
  A_3(1^{h_1}2^{h_2}3^{h_3})=c' &\omega^{-h_1}_1\omega^{-h_2}_2\omega^{-h_3}_3z^{h_3-h_1-h_2}_{12}z^{h_2-h_1-h_3}_{31}z^{h_1-h_2-h_3}_{23}\delta^{(3)}(\sum_ip_i)\,,
\end{align}
and similarly for the barred sector.
In order for this to be finite, one has to resolve the delta function on the barred sector.
For comparison with the main text let us
write this in position space using the map presented in
\cite{Donnay:2022wvx}. We find
\begin{align}
  \label{eq:18a}
  \mathcal{C}_3=c''\prod^3_{i=1}\int^\infty_0 &\frac{\dd\omega_i}{\omega^{h_i}_i} e^{-i\veps_i\omega_iu_i}z^{h_3-h_1-h_2}_{12}
z^{h_2-h_1-h_3}_{13}z^{h_1-h_2-h_3}_{23} \delta^{(4)}(\sum_i\veps_i\omega_iq_i)\,.
\end{align}
with $c''=c'/(2\pi)^3$.
Following \cite{Pasterski:2017ylz}, we introduce simplex variables $\omega_i=t_i\omega$ with $\sum t_i=1$ and $\sum_i\omega_i=\omega$ and resolve the momentum conserving delta functions on the barred sector (cf., e.g., \cite{Pate:2019mfs}) to find
\begin{align}
  \label{eq:18}
  \mathcal{C}_3=c''\int^\infty_0\frac{\dd\omega}{\omega^{2+h_1+h_2+h_3}}
                 \int^1_0\frac{\dd t_i}{t^{h_i}_i}&
                                                    e^{-i\omega\veps_i t_i u_i} z^{h_3-h_1-h_2}_{12} z^{h_2-h_1-h_3}_{13} z^{h_1-h_2-h_3}_{23}\nonumber\\
  \frac{1}{t_1t_2t_3D^2_3}
                &\delta\left(t_1-\frac{z_{23}}{D_3}\right)
                 \left(t_2-\frac{\veps_1\veps_2 z_{31}}{D_3}\right)
                 \left(t_3-\frac{\veps_3\veps_1 z_{12}}{D_3}\right)
                 \delta(\zb_{12})\delta(\zb_{13})\,
\end{align}
where $D_3$ is the Jacobian
\begin{equation}
  \label{eq:1}
  D_3=(\veps_1\veps_2-1)z_{31}+z_{12}(\veps_1\veps_3-1)\,.
\end{equation}
Due to the delta functions all integrals over the simplex variables $t_i$ collapse to their values $t^*_i$ and we are left with
\begin{align}
  \label{eq:6} 
\mathcal{C}_3&=(\veps_1\veps_2)^{h_2+1}(\veps_1\veps_3)^{h_3+1}\int^\infty_0\frac{\dd\omega}{\omega^{2+h_1+h_2+h_3}}  e^{-i\veps_1\frac{\omega}{z_{23}}(u_1z_{23}+u_2z_{31}+u_3z_{12})})\nonumber\\ &\qquad\qquad z^{-1-h_1-h_2}_{12}z^{-1-h_1-h_3}_{13}z^{h_1}_{23}\delta(\zb_{13})\delta(\zb_{12})
  \prod^3_{i=1}\mathbf{1}_{[0,1]}(t^*_i),
\end{align}
where we rescaled the integration variable by $D_3/z_{23}$, which is always positive.
Here $\mathbf{1}_{[0,1]}(t^*_i)$ is an indicator function that equals to one when its argument is in the interval $[0,1]$. The indicator function can be written equivalently as
\begin{equation}
  \label{eq:25}
  \prod^3_{i=1}\mathbf{1}_{[0,1]}(t^*_i)=\Theta\left(\frac{\veps_1 z_{23}}{\veps_2z_{31}}\right)\Theta\left(\frac{\veps_1 z_{23}}{\veps_3 z_{12}}\right)\,.
\end{equation}
This allows us to write \eqref{eq:6} as
\begin{align}
  \label{eq:43}
  \mathcal{C}_3=c''&\sgn(z_{12}z_{23}z_{31})^{|h_1|+|h_2|+|h_3|}\int^\infty_0\frac{\dd\omega}{\omega^{2+h_1+h_2+h_3}}  e^{-i\veps_1\frac{\omega}{z_{23}}(u_1z_{23}+u_2z_{31}+u_3z_{12})}\nonumber\\
  &|z_{12}|^{-1-h_1-h_2}|z_{13}|^{-1-h_1-h_3}|z_{23}|^{h_1}\delta(\zb_{13})\delta(\zb_{12})
  \prod^3_{i=1}\mathbf{1}_{[0,1]}(t^*_i),
\end{align}
where we assumed that all $h_i$ are integer. Rotating the contour slightly into an imaginary direction by adding $-\epsilon \omega $, introducing g a deformation parameter $\Delta$ and setting $h_1+h_2+h_3\equiv h$, let us define
\begin{align}
  \label{eq:86}
  &\int^\infty_0\frac{\dd\omega}{\omega^{2+h}}  e^{-i\veps_1\frac{\omega}{z_{23}}(u_1z_{23}+u_2z_{31}+u_3z_{12})}=
  \lim_{\Delta\to 3,\epsilon\to 0} \int^\infty_0\frac{\dd\omega}{\omega^{5+h-\Delta}}  e^{-i\veps_1\frac{\omega}{z_{23}}(u_1z_{23}+u_2z_{31}+u_3z_{12})-\epsilon\omega}\\&=\lim_{\Delta\to 3,\epsilon\to 0} \frac{i^{\Delta-h-4}\Gamma(\Delta-h-4)}{(\frac{\veps_1}{z_{23}}(u_1z_{23}+u_2z_{31}+u_3z_{12})-i\epsilon)^{\Delta-h-4}}.\nonumber
\end{align}
The three-point function in position space becomes then
\begin{align}
  \label{eq:threepointinpos}
\mathcal{C}_3=c''&\sgn(z_{12}z_{23}z_{31})^{|h_1|+|h_2|+|h_3|}|z_{12}|^{-1-h_1-h_2}|z_{13}|^{-1-h_1-h_3}|z_{23}|^{h_1}\delta(\zb_{13})\delta(\zb_{12})
  \prod^3_{i=1}\mathbf{1}_{[0,1]}(t^*_i)\nonumber\\&\lim_{\Delta\to 3,\epsilon\to 0} \frac{i^{\Delta-h-4}\Gamma(\Delta-h-4)}{(\frac{\veps_1}{z_{23}}(u_1z_{23}+u_2z_{31}+u_3z_{12})-i\epsilon)^{\Delta-h-4}}.
\end{align}
The three-point correlators computed in the main text are precisely of this form.
Note that the corresponding correlators for the ``news fields'', obtained by three partial derivatives w.r.t. $u_1,u_2,u_3$ are all finite.

\providecommand{\href}[2]{#2}\begingroup\raggedright\endgroup

\end{document}